\documentclass[%
 reprint,
 amsmath,amssymb,
 aps,
pra,
]{revtex4-1}

\usepackage{graphicx}
\usepackage{dcolumn}

\usepackage{amssymb}
\usepackage{amsmath}
\usepackage[utf8]{inputenc}
\newcommand\dif{\mathop{}\!\mathrm{d}}

\usepackage{bm}
\usepackage[dvipsnames]{xcolor}  
\definecolor{darkgreen}{rgb}{0.0, 0.5, 0.0}

\usepackage[english]{cleveref}
\usepackage{url}
\begin{document}

\preprint{}

\title{Electronic transport through correlated electron systems\\ with nonhomogeneous charge orderings}

 \author{Rudolf Smorka}
 \affiliation{Institute of Physics, Albert-Ludwigs-Universität Freiburg, Hermann-Herder-Straße
3, 79104 Freiburg i. Br., Germany}
 
 \author{Martin \v{Z}onda}
 \email{martin.zonda@physik.uni-freiburg.de}
 \affiliation{Institute of Physics, Albert-Ludwigs-Universität Freiburg, Hermann-Herder-Straße
3, 79104 Freiburg i. Br., Germany}
 
 \author{Michael Thoss}
 \affiliation{Institute of Physics, Albert-Ludwigs-Universität Freiburg, Hermann-Herder-Straße
3, 79104 Freiburg i. Br., Germany}

\date{\today}

\begin{abstract}
The spinless Falicov-Kimball model exhibits outside the particle-hole symmetric point different stable nonhomogeneous charge orderings. These include the well known charge stripes 
and a variety of orderings with phase separated domains, which can significantly influence the charge transport through the correlated electron system.
We show this by investigating a heterostructure, in which the Falicov-Kimball model on a finite two-dimensional lattice is located between two noninteracting semi-infinite leads. We use a combination of nonequilibrium Green's functions techniques with a sign-problem-free Monte Carlo method for finite temperatures 
or a simulated annealing technique for the ground state to address steady-state transport through the system. We show that different ground-state phases of the central system can lead to simple metallic-like or insulating charge transport characteristics, but also to more complicated current-voltage dependencies reflecting a multi-band character of the transmission function. Interestingly, with increasing temperature, the orderings tend to form transient phases before the system reaches the disordered phase. This leads to nontrivial temperature dependencies of the transmission function and charge current.

\end{abstract}

\maketitle

\section{Introduction}

Nonhomogeneous long-range charge and spin orderings have been 
observed in a multitude of strongly correlated electron systems.
Among the most studied are the static and fluctuating stripe-like 
charge orderings, observed in doped copper oxides, that
interfere with the high-temperature superconductivity \cite{Tranquada95,Berg2009,Parker2010,WuTao2011,Abbamonte2005,Tranquada2008,Fausti2011,Ghiringhelli2012,Comin2015,Jacobsen2018,Zhao2019}.
Equivalent structures have been observed 
in a variety of other materials including layered cobalt oxides \cite{Boothroyd2011,Babkevich2016} 
and nickelates \cite{Poltavets2010,Coslovich2017,Zhang2017,Norman2016,Zhang2019}. 
The nonhomogeneous long-range orderings are often accompanied by a phase separation 
characterized by domains of mutually different phases \cite{Tranquada1999,Julien1999,Coslovich2017}, and this tendency toward the electronic phase separation and 
pattern formation seems to be rather general in strongly correlated electron systems \cite{Neto2004,Daggoto2001,Tokura2006,Vojta2009}.

Different theoretical approaches have been applied to describe the formation
of nonhomogeneous orderings in these materials. 
Important insight into the problem was gained by 
phenomenological models \cite{Eskes1996,Eskes1998,Kivelson1998,Dimashko1999,Bogner2001,Vojta2006,Vojta2009}
and in works focusing on the competition between 
the long-range interactions and the natural tendency of 
strongly correlated electron systems to phase separation \cite{Emery1990,Pryadko1998,Pryadko1999}.   
However, it was also shown that various simplified models of correlated electrons 
that consider only local and nearest neighbor interactions 
can naturally describe the formation of nonhomogeneous ground states including charge and spin stripes \cite{White1998,White1998b,Hotta2000,LemanskiPRL2002,Himeda2002,Hager2005,Chang2010,Cencarikova2011,Corboz2014,Yamase2016,Zheng2017,KapciaPRE2017,Huang2018,Jiang2019}.

The simplest of these models is the Falicov-Kimball model (FKM) \cite{FKM1969} 
in which the itinerant electrons interact with localized particles.
The ground state of the spinless FKM at the particle-hole symmetric (PHS) point is a charge density wave (CDW) phase for any finite Coulomb interaction on a bipartite lattice in dimension two or higher \cite{Brandt1989,Brandt1990,Brandt1991} and this phase is stable up to finite critical temperatures 
\cite{FreericksDMFT2003,ChenPRB2003,MaskaPRB2006,Kapcia2019,ZondaSSC2009}.
However, it was also proven that a segregated phase, which is a special case of phase separation 
where the heavy particles (ions) and electrons reside in separate domains, 
is the ground state in the limit of infinite Coulomb interaction of the spinless version of the FKM 
\cite{Freericks1990,Lemberger1992,Letfulov1998,Freericks1999,Freericks2000,Freericks2002}. 
Away from these special limits, hence, for moderate Coulomb interaction and outside the PHS point, the ground state of the FKM
is extremely rich and contains stable
nonhomogeneous charge orderings of various types including different types of charge stripes \cite{Lemanski1995,LemanskiPRL2002,Lemanski2004,Farkasovsky2005,Brydon2006,FarkyAPS2010} which are stable also at finite temperatures \cite{Tran2006,Czajka2007,MaskaPRA2011,Zonda2012,Debski2016,Zonda2019}.
Besides, these already exotic charge orderings are often accompanied by nonhomogeneous magnetic structures
in the spinful version of the FKM and its various generalizations
\cite{Lemanski2005,Cencarikova2008,ZondaPhT2009,Cencarikova2011}.

It is worth emphasizing that nonhomogeneous charge phases play an important role in 
studies of various phenomena modeled by the FKM including crystallization \cite{Kennedy_1986,Jedrzejewski1989,Gruber1994}, metal-insulator and valence transitions \cite{Plischke1972,Michielsen1994,Farky1995a,Farky1995b,Portengen1996,Czycholl1999,Byczuk2005,Farky2019,Nasu2019,Haldar2019}, localization \cite{Byczuk2005,Maionchi2008,Carvalho2014,Haldar2017b,Zonda2019b}, distribution of heavy and light cold
atoms in optical lattices \cite{MaskaPRL2008,Iskin2009,MaskaPRA2011,HuMaska2015,QinPRA2018}, or studies of nonlocal correlations \cite{Schiller1999,Hetler2000,Ribic2016,Ribic2017}. 
They should be also considered when
addressing different nonequilibrium phenomena \cite{FreericksPRL2006,Turkowski2007,EcksteinPRL2008,Eckstein2009,Matveev2016,Herrmann2016,Haldar2016,Smith2017,Herrmann2018} and any type of transport \cite{FreericksBook2006,MatveevPRB2008,Okamoto2007,Okamoto2008,Haldar2017,Zonda2019,Zonda2019b} while dealing with a system outside the PHS point, e.g., a doped system. 

\begin{figure}[!h]
   \centering
    \includegraphics[width=1.00     \columnwidth]{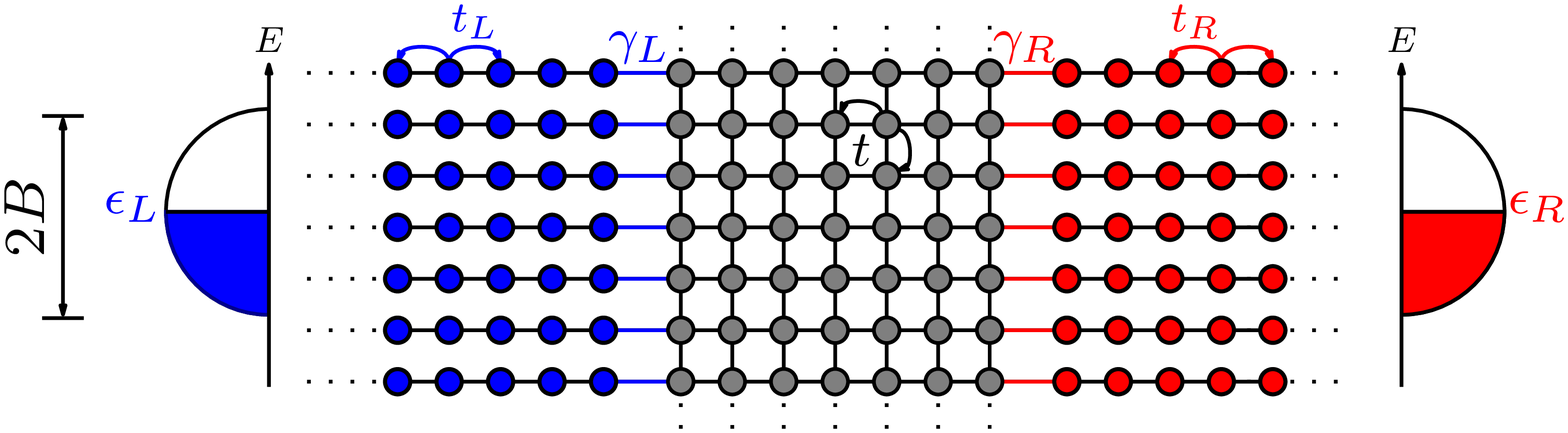}
   \caption{Heterostructure with the central spinless FKM system sandwiched between two noninteracting semi-infinite metallic leads with semielliptical surface DOS of total width $2B$ centered around $\epsilon_{L,R}$. The nearest-neighbor hopping is set by the matrix element $t$ which we use as the unit of energy. The exchange of the $d$ electrons between the system and leads is given by the couplings $\gamma_{L}=\gamma_{R}=2t$ and the hopping in the leads by $t_{l}=B/2=10t$. \label{fig:schema}}
   
   \label{fig:Heterostructure}
\end{figure}

In our work we are especially interested in nonequilibrium steady-state charge transport through a finite layered FKM system (for the sake of clarity, we refer by ``system" just to the central part of the heterostructure shown in Fig.~\ref{fig:schema}) sandwiched between two semi-infinite metallic leads. 
This, or similar setups, has been addressed before \cite{Zonda2019, FreericksBook2006, Okamoto2008, Zonda2019b}. However, the regimes of nonhomogeneous charge orderings, such as stripes or phase separation,
have not been considered. 

We focus on the regime of intermediate Coulomb interaction 
between the constituents for which it was shown in previous works 
that it is rich in nonhomogeneous charge orderings, which are stable 
even at finite temperatures \cite{Lemanski1995,LemanskiPRL2002,Lemanski2004,Tran2006,Zonda2012}. 
Here, we address transport through a system that is in principle infinite in the 
direction parallel to the system-leads interfaces ($y$), 
but finite in the perpendicular direction ($x$) as illustrated in Fig.~\ref{fig:schema}. 
This allows us to investigate the dependence of the 
transport on the width of the system starting with a lattice just a few layers thick. 
To model this we use a central system with mixed boundary conditions, namely, open ones 
at the system-leads interfaces and periodic in the other direction.

As far as we know, this system geometry has not been addressed in the literature away from PHS point. 
Therefore, building on the prior studies of orderings in the FKM on regular
square lattices with periodic boundary conditions \cite{Lemanski1995,LemanskiPRL2002,Lemanski2004,Tran2006,Zonda2012},
we first investigate the stability of the nonhomogeneous charge orderings in our setup.
We focus on the influence of particle filling, the thickness of the 
system, and also temperature on the charge ordering in the system. 
We then show how various homogeneous configurations and their change influence the 
nonequilibrium charge transport properties.

The rest of the paper is organized as follows. In Sec.~\ref{MaM}, we introduce the model and outline the methodology to characterize the ground-state and finite-temperature properties. 
In Sec.~\ref{ZeroT} we focus on transport through the ground-state orderings of various particle concentrations. 
In Sec.~\ref{FinT} we study finite-temperature transport properties for representative cases that evolve with increasing temperature from the low-temperature nonhomogeneous phases through a transient ordering to a high-temperature disordered phase.   
Section~\ref{Summary} concludes with a summary. 
In Appendix~A we address the stability of the 
density of electrons for varying electrochemical potential, voltage, or temperature; 
in Appendix~B we discuss the finite-size scaling of some transport properties; and Appendix~C is dedicated to the
influence of different types of particle excitations on the charge current.

\section{Model and methods \label{MaM}}

\subsection{Model}
We consider a heterostructure consisting of a central system of correlated electrons on a lattice sandwiched between two noninteracting leads as illustrated in Fig.~\ref{fig:schema}.   
The physics of the central system, where the itinerant $d$ electrons and localized fermionic $f$ particles 
interact via local Coulomb interaction, is described by the spinless Falicov-Kimball Hamiltonian \cite{FKM1969,Kennedy_1986}
\begin{align}
H_\mathrm{s} = & -\sum_{\langle j,j'\rangle} t_{jj'}d_{j}^{{\dagger}} d_{j'} + U\sum_j d_{j}^{\dagger} d_{j} f_{j}^{\dagger} f_j \notag \\ & - \sum_j \overline{\mu}_j d_{j}^{\dagger} d_j. 
\label{eq:CentralHamiltonian}
\end{align}
Here, the first term describes quantum mechanical tunneling of $d$ electrons from site $j$ to $j'$ 
on a lattice. 
The lattice geometry and boundary conditions are set by the hopping matrix with the elements $t_{jj'}$. 
We assume only hopping to the nearest-neighbor sites with a constant amplitude $t$ which sets the unit of energy in our system.
We consider a square lattice central system in two dimensions, typically with number of layers $L_x$ and number of sites per layer $L_y$ being unequal. 
We employ mixed system boundary conditions, namely, open at the edges coupled to the leads and periodic boundary conditions in the $y$ direction. 

The second term in Eq.~\eqref{eq:CentralHamiltonian} describes local Coulomb-like interaction between the $d$ electrons and localized $f$ particles.
With the last term, the position-dependent electrochemical potential $\overline{\mu}_j$ is taken into account,
combining both the electrostatic and chemical potential of the decoupled system, acting on $d$ electrons. 
In principle, the profile of this potential can be influenced by the leads \cite{FreericksBook2006,Okamoto2007,Okamoto2008,Gaury2014,Zonda2019} acting as
reservoirs for the itinerant $d$ electrons. 
In our study, we assume that the electrochemical potential is constant in the whole system 
$\overline{\mu}_j=\overline{\mu}$. 
The influence of the leads is taken into account implicitly as a constant shift of this potential. 
A similar term for the $f$ particles is neglected as we are keeping their concentration fixed because 
their number is not altered by the leads.

The central system is sandwiched between two metallic leads, which are modeled by 
\begin{equation}
    H_{\mathrm{lead}}^{l}=-t_l\sum_{\left\langle n,n' \right\rangle }\left(c_{l,n}^{\dagger}c_{l,n'}+c_{l,n'}^{\dagger}c_{l,n}\right)+\epsilon_{l}\sum_{n}c_{l,n}^{\dagger}c_{l,n}, \label{eq:Lead}
\end{equation}
where $t_{l}$ is the hopping constant for lead $l=L,R$ and $\epsilon_l$ represents
a constant energy shift of the respective lead. 
The coupling between the central system and leads is described by the hybridization term
\begin{equation}
H_{\mathrm{hyb}}^{l}  =  -\gamma_{l}\sum_{\left\langle i,n\right\rangle }\left(c_{l,n}^{\dagger}d_{i}^{\phantom{\dagger}}+d_{i}^{\dagger}c_{l,n}^{\phantom{\dagger}}\right),\label{eq:hyb}
\end{equation}
with $\gamma_{l}$ being the lead-system coupling.

We assume that the semi-infinite leads are
unaffected by the system and model them by parallel chains coupled
to the central system as shown in Fig.~\ref{fig:schema}. 
Therefore, the leads can be characterized by their surface density of states (DOS)
\begin{equation}
\rho_{l}\left(E\right)=\frac{2}{\pi B^{2}}\sqrt{B^{2}-\left(E-\epsilon_{l}\right)^{2}},
\end{equation}
with half-bandwidth $B=2t_{l}$ centered around the band energy shift
$\epsilon_{l}$ from Eq.~\eqref{eq:Lead} \cite{Cizek2004,Zonda2019}. We keep the leads half filled by 
fixing $\mu_l=\epsilon_l$, where $\mu_l$ is the chemical potential of the leads. 
The voltage drop $V=\mu_L-\mu_R$ is then introduced by mutually shifting $\mu_L$ and $\mu_R$.

\subsection{Methods}

We are utilizing a combination of a sign-problem-free Monte Carlo (MC) method \cite{MaskaPRB2006,ZondaSSC2009,Zonda2012,Antipov2016,Huang2017} with a nonequilibrium
Green’s function technique \cite{QKinetics,Spicka2014}, which allows us to address
nonequilibrium steady-state charge transport in the heterostructure \cite{Zonda2019,Zonda2019b}. 
This method takes advantage of the fact that the occupation 
numbers $n^f_{j}=f^\dagger_jf^{\phantom{\dagger}}_j$ of the localized $f$ particles are integrals of motion and, therefore, 
their distribution does not change in time.
In addition, there is no reservoir that would allow for a fast
thermal rearrangement of the $f$ particles after the system is coupled to the leads. 
This describes a situation, in which the $f$ particles are effectively so much heavier than the
conducting $d$ electrons that the 
investigated steady state for the $d$ subsystem 
is reached much faster than
any non-equilibrium effect on the $f$-particle distribution can be observed 
\footnote{The realistic time scale of a potential rearrangement of the $f$-particle in the presence of
an additional reservoir could be, in principle, estimated by kinetic MC or equivalent methods.}. 
In our study, we assume that in the distant past the system was decoupled from the leads
and both the system and the leads had been in thermal equilibrium.
This means that the steady state depends on the 
equilibrium $f$-particle distribution \cite{Okamoto2007,EcksteinPRL2008,Zonda2019}.

The value of electrochemical potential, which governs the number of $d$ electrons in the decoupled system,
is dictated by the equilibrium state of the semi-infinite reservoirs ($\mu_L=\mu_R$) before the leads are coupled. 
Formally, we take $\overline{\mu}$ to be part of the system Hamiltonian. 
This choice shifts the system Fermi level to zero energy and makes 
the voltage drop in the leads symmetrical around zero ($\mu_L=-\mu_R$). 

If we split the systems electrochemical potential into its constituents,
another interpretation of our protocol is possible.
Namely, that the chemical potential of the decoupled system is always zero and
that the number of $d$ electrons, and consequently the $f$-particle distribution,
are both dictated by the varying flat electrostatic potential, as is common in 
studies of pressure-induced valence and metal-insulator transitions \cite{Farky1995a,Farky1995b,Farky2019,Nasu2019}. 
No matter the interpretation, the method requires us to first 
investigate the thermal distribution or the ground-state $f$-particle configurations of the central system.   

\subsubsection{Decoupled system}

The fact that the $f$-particle number operators $n^f_j$ are good quantum numbers 
can be used to simplify the system Hamiltonian in Eq.~\eqref{eq:CentralHamiltonian} by replacing $n^f_j$ with its eigenvalues $w_j=1$ (occupied) or $0$ (unoccupied). 
The simplified Hamiltonian for a fixed $f$-particle configuration $w$ thus reads 
\begin{equation}
H^{w}_{\mathrm{S}} =  \sum_{j,j'} h_{jj'}d_{j}^{{\dagger}} d_{j'} 
= \sum_\alpha \lambda_\alpha b_\alpha^\dagger b_\alpha \label{eq:CentralHamiltonianSimpl}
\end{equation}
with $h_{jj'}=(Uw_j-\overline{\mu})\delta_{jj'}-t_{jj'}$. 
Here, a unitary transformation  $\bm{\lambda}=\bm{\mathcal{U}} \bm{h}\bm{\mathcal{U}}^\dagger$ is used to diagonalize the system matrix $\bm{h}$ with $\lambda_\alpha$ being its eigenvalues stored in ascending order. 
Finding the ground state then means to find the configuration $w$ with the lowest energy, 
\begin{equation}
    E^{\mathrm{gs}}_{\mathrm{S}}(w)=\sum_{\alpha=1}^{L}\lambda^w_\alpha \Theta(-\lambda^w_\alpha),  \label{eq:Egc}
\end{equation}
where $\Theta(E)$ is the Heaviside step function.
Although this is formally simple, finding the ground-state configuration for a large lattice size $L=L_x \times L_y$  is computationally demanding, often out of reach for the current state computational techniques. 
Therefore, we determine the ground state approximately using a simulated annealing method \cite{Cerny1985} built on
the same MC technique which we use for finite temperatures.

The MC method takes advantage of the fact that the mean values
of any local $d$-electron operator $\hat{O}$ of the FKM can be written in the form
\begin{equation}
\left\langle \hat{O}\right\rangle =\mathrm{Tr}_{w}\left\langle \hat{O}\right\rangle _{d}\equiv\frac{1}{Z}\sum_{w}e^{-\beta F\left(w\right)}\left\langle \hat{O}\right\rangle _{d},\label{eq:Aver}
\end{equation}
where 
\begin{equation}
F\left(w\right)=-\frac{1}{\beta}\sum_{\alpha}\ln\left[1+e^{-\beta\lambda^{w}_{\alpha}}\right],\label{eq:Free}
\end{equation}
with $Z=\sum_{w}e^{-\beta F\left(w\right)}$ being the partition function
(remember that $\overline{\mu}$ is incorporated into the Hamiltonian). 
Here, $\left\langle .\right\rangle _{d}$ is
the trace over the $d$-electron subsystem for fixed $w$ \cite{MaskaPRB2006}.
As this is a single particle problem, the trace can be calculated
efficiently using exact numerical diagonalization. The sum over configurations
$w$ can then be calculated using a Metropolis-algorithm-based MC
method \cite{MaskaPRB2006,ZondaSSC2009,ZondaPhT2009,Maska2005,Zonda2012,Antipov2016,Huang2017}.

The simulated annealing method \cite{Cerny1985}, which we use for finding the ground-state configurations, is methodically equivalent to the MC method, where we start at a high temperature and decrease it steadily to zero. 
However, instead of the free energy in Eq.~\eqref{eq:Free} we use as weights the 
ground-state energy $E^{\mathrm{gs}}_{\mathrm{S}}(w)$ from Eq.~\eqref{eq:Egc}, which allows us to reach lower temperatures.
In some cases, the temperature at which the ground-state configuration starts to order 
is very low, making the annealing process inefficient. 
For these cases we have also applied a simple variation of the zero-temperature hill climbing algorithm described in Ref.~\cite{Farky2001}.   

Besides the distribution of $f$ particles, we use other properties of the isolated system to explain the character of charge transport. 
In this context, the most important property is the system density of states 
DOS$(E)$=Tr$_w$DOS$(E,w)$ with 
 \begin{align}
     \mathrm{DOS}(E,w)&=\frac{1}{L}\sum_\alpha \delta(E-\lambda^w_\alpha) \notag\\ &\approx\frac{1}{\sqrt{4\pi\sigma}L}\sum_{\alpha} \mathrm{e}^{-\frac{(E-\lambda^w_\alpha)^2}{4\sigma}},
     \label{eq:DOS}
 \end{align}
where we use a Gaussian broadening of otherwise sharp states $\lambda^\omega_\alpha$ with $\sigma=0.1t$ when addressing the ground-state properties and $\sigma=0.02t$ otherwise. This artificial broadening is used only to smooth the data in 
the figures. It has no effect on the transport in the heterostructure 
because there only the broadening which arises naturally 
from the coupling to the leads is taken into account.

We also utilize the specific heat, defined as
\begin{equation}
C_V = \frac{\beta^2}{L}\left(\left<E^2\right> -\left<E\right>^2 + \left<\sum_\alpha \frac{(\lambda^w_\alpha)^2}{e^{\beta\lambda^w_\alpha}+e^{-\beta\lambda^w_\alpha}+2}\right>\right),
\end{equation}
and various order parameters to identify the approximate temperatures below which nonhomogeneous orderings start to form.

\subsubsection{Coupled system}

The transformed Hamiltonian in Eq.~\eqref{eq:CentralHamiltonianSimpl} describes,
for a specific configuration $w$, a noninteracting system.  
Nonequilibrium transport in this kind of model is a well-studied 
problem \cite{Jauho1994,Cizek2004,Zonda2019} 
and the exact form of the steady-state Green's functions is given by \cite{QKinetics} 
\begin{eqnarray}
\mathbf{G}^{r,a}\left(E\right) & = & \bm{g}^{r,a}\left(E\right)+\bm{g}^{r,a}\left(E\right)\mathbf{\Sigma}^{r,a}\left(E\right)\mathbf{G}^{r,a}\left(E\right),\label{eq:Gra}\\
\mathbf{G}^{<}\left(E\right) & = & \mathbf{G}^{r}\left(E\right)\mathbf{\Sigma}^{<}\left(E\right)\mathbf{G}^{a}\left(E\right).
\end{eqnarray}
Here, $\mathbf{G}^{r\,(a)}$ is the retarded (advanced) Green's function
of the coupled system, $\mathbf{G}^{<}$ is the lesser Green's function,
and $\bm{g}^{r\,(a)}\left(E\right)$ is the retarded
(advanced) Green's function of the bare system with components 
\begin{equation}
g_{\alpha\beta}^{r,a}\left(E\right)=\frac{\delta_{\alpha\beta}}{E-\lambda^w_{\alpha}\pm i0}.
\end{equation}
The total tunneling self-energies of the simplified leads
$\mathbf{\Sigma}^{r,a,<}=\mathbf{\Sigma}_{L}^{r,a,<}+\mathbf{\Sigma}_{R}^{r,a,<}$
have the components 
\begin{eqnarray}
\Sigma_{l,\alpha\beta}^{r,a}(E) & = & \Lambda_{l,\alpha\beta}(E)\pm\frac{i}{2}\Gamma_{l,\alpha\beta}(E),\nonumber \\
\Sigma_{l,\alpha\beta}^{<}(E) & = & i\Gamma_{l,\alpha\beta}(E)\,f_{l}(E-\mu_{l}),\nonumber \\
\Gamma_{l,\alpha\beta}(E) & = & 2\pi\gamma^{2}\mathcal{U}_{\alpha\beta}^{\{s^{l}\}}\rho_{l}(E),\label{eq:Gamma}\\
\Lambda_{l,\alpha\beta}(E) & = & \begin{cases}
\frac{2\gamma^{2}}{B^{2}}\mathcal{U}_{\alpha\beta}^{\{s^{l}\}}\left(E-\epsilon_{l}\right)\:\textrm{for}\:\left|E-\epsilon_{l}\right|<B\\
\frac{2\gamma^{2}}{B^{2}}\mathcal{U}_{\alpha\beta}^{\{s^{l}\}}\left[\left(E-\epsilon_{l}\right)\mp\sqrt{\left(E-\epsilon_{l}\right)^{2}-B^{2}}\right]\\
\qquad\textrm{for}\:\left(E-\epsilon_{l}\right)\gtrless\pm B
\end{cases}\nonumber \\
\mathcal{U}_{\alpha\beta}^{\{s^{l}\}} & = & \sum_{\begin{array}{c}
i\in\{s^{l}\}\end{array}}\mathcal{U}_{\beta i}^{\dagger}\mathcal{U}_{i\alpha},\nonumber 
\end{eqnarray}
where $\{s^{L,R}\}$ are the sets of system lattice positions at the
left and right interfaces and $f_{l}(E)$ is the Fermi function. 
This exact analytical form of the self-energies 
follows from the fact that leads in our model consist of independent semi-infinite chains, each connected to the system only through one site. The details of the derivation of Eq.~\eqref{eq:Gamma} can be found in Refs.~\cite{Economou2006, Cizek2004, Zonda2019}.

Using the standard Landauer-B\"{u}ttiker formula, the steady-state current $I^w$  for a 
specific $f$-particle configuration is given by  
\begin{equation}
    I^w=\int \frac{\dif E}{2\pi}T(E,w)[f(E-\mu_L)-f(E-\mu_R)],
\end{equation}
where $T(E,w)$ is the transmission function given by the trace over the $d$-electron subsystem:
\begin{equation}
    T(E,w)=\mathrm{Tr}_d\{\bm{\Gamma}_L(E)\bm{G}^r(E)\bm{\Gamma}_R(E)\bm{G}^a(E)\}.
\end{equation}
Other quantities that we are interested in are the spatially 
resolved nonequilibrium  densities of the conducting $d$ electrons,
\begin{equation}
\left<n_{d,j}^\text{NE}\right>_d=-i\int dE \sum_{\alpha,\beta}\mathcal{U}_{j\alpha} \mathcal{U}_{\beta j}^{\dagger} G_{\alpha \beta}^<(E)   
\end{equation} 
and the local DOS of the coupled system,
\begin{eqnarray} 
\mathrm{LDOS_{\mathit{j}}}(E,w) & = & \frac{i}{2\pi L}\left[\sum_{\alpha,\beta}\mathcal{U}_{j\alpha}\mathcal{U}_{\beta j}^{\dagger}G_{\alpha\beta}^{r}\left(E\right)\right.\label{eq:LDOS}\\
 &  & \left.-\sum_{\alpha,\beta}\mathcal{U}_{\alpha j}^{\dagger}\mathcal{U}_{j\beta}G_{\alpha\beta}^{a}\left(E\right)\right]\nonumber,
\end{eqnarray}
which allows us to define the generalized inverse participation
ratio (gIPR) 
\begin{equation}
\mathrm{gIPR}(E,w)=\frac{\sum_{j}\mathrm{LDOS_{\mathit{j}}^{2}\mathit{(E,w)}}}{\left[\sum_{j}\mathrm{LDOS_{j}\mathit{(E,w)}}\right]^2}.\label{eq:gIPR}
\end{equation}
The inverse participation ratio and its
generalization can be used to identify the localization of itinerant
electrons in strongly correlated electron systems \cite{Evers2008,Murphy2011,Perera2018,Antipov2016}.
The gIPR scales as $1/L$ for completely itinerant
states, as $\sim 1/L_y$ for energies within the gap and outside the
central system DOS, where the dominant contribution
comes from the LDOS at the interfaces, and it converges to a
finite value with increasing $L$ for localized states.

Because the $f$-particle occupation numbers are integrals of motions, 
the thermal mean values of above quantities can be evaluated by
employing the same MC procedure as introduced for the decoupled system~\cite{Zonda2019}.

\section{Results}

It was shown in previous studies 
that the FKM with intermediate interaction $U$
exhibits a vast variety of stable
charge configurations \cite{LemanskiPRL2002,Lemanski2004,Tran2006,MaskaPRA2011,Zonda2012}.
Here, we address a few typical, but nontrivial, examples of this regime
represented by $U=4t$. The stability of nonhomogeneous phases for $U=4t$ was investigated in detail in several studies; therefore, various approximate phase diagrams are known \cite{Lemanski2004,Czajka2007,Zonda2012}. Although
these diagrams have been calculated for different system geometry and boundary conditions, we use them as a guide for setting the concentration of the $f$ particles
which heavily influences the charge ordering in the system.  
The system ground-state concentrations of the $d$ electrons ($n_d$)
are set by the electrochemical potential (for details see Appendix~A). 
We focus on two types of fillings studied in the literature, 
namely the neutral (or nearly neutral) case
$n_f\simeq n_d$ \cite{LemanskiPRL2002} and the half-filled case $n_f+n_d\simeq 1$.
Note that the similarity signs refer 
to the fact that we fix the same $\overline{\mu}$ for all temperatures, lattice sizes and voltages.
Under this condition, the total density of $d$ electrons is stable with changing lattice size and reasonably 
stable with the temperature. However, it varies with the voltage because of the
asymmetric DOS of the $d$ electrons.
We address this problem in detail in Appendix~A.  
 
In our analysis, we first focus on charge transport at zero temperature because this allows us to discuss 
the direct influence of the specific $f$-particle orderings without any thermal fluctuations.
In the following we use leads with half-bandwidth $B=20t$, which was chosen 
to be broad enough to incorporate the 
whole DOS of the isolated system when $V\ll B$. 
 We
set $\gamma=\gamma_{L}=\gamma_{R}=2t$, which corresponds only to a fifth of the
hopping amplitude in the leads (which follows Ref.~\cite{Cizek2004}), but still provides 
a sufficient broadening of the system states.
Note that the charge-current dependence on $\gamma$ is, in general, non-linear and 
might be nonmonotonic even in the simple case of a noninteracting single resonant 
level model \cite{QKinetics}. However, our tests showed that the choice
$\gamma=2t$ is at a range of values, where charge current increases with $\gamma$ approximately  
linearly for all voltages.

For every investigated case we first address the ground-state real-space $f$-particle distribution.
We show how it depends on system width
by considering three lattice sizes,
$L=L_x\times L_y=6\times 24, 12\times 24$ and $24\times 24$. 
We use $L_y=24$ because it was shown recently for the PHS point that $L_y\sim 20$ is sufficient to approximate a layered system with $L_y\rightarrow\infty$ \cite{Zonda2019} and similar conclusions can be drawn outside this point as briefly discussed in Appendix~B. 
Afterward, we address the related DOS of the $d$ electrons for the isolated system.  
The ground-state orderings of $f$ particles and the DOS of the $d$ electrons are often sufficient 
to explain the main qualitative features of the $I-V$ characteristics and transmission functions 
that give detailed information about the nonequilibrium charge transport.
In cases where deeper analysis is needed
we discuss also other quantities such as the gIPR or order parameters.
Note that because we are always addressing transport in a finite system
we call an $I-V$ characteristic insulating, if it clearly reflects a large energy gap
on the Fermi level with negligible current for small voltages, and 
we call it metallic-like if
the current increases approximately linearly already for small voltages.  

\subsection{Zero temperature \label{ZeroT}}
\subsubsection{Particle-hole symmetric case \label{PHSC}}
We start our analysis by discussing the PHS case $\overline{\mu}=U/2$ which sets $n_f=n_d=0.5$ 
and is, therefore, both neutral and half-filled. 
This regime was investigated in detail in previous works \cite{Zonda2019,Zonda2019b}, 
but it is useful to recapitulate the main results here as we will later put them into contrast with the situation 
away from the PHS point.

\begin{figure}[!h]
    \centering
    \includegraphics[width=0.95\linewidth]{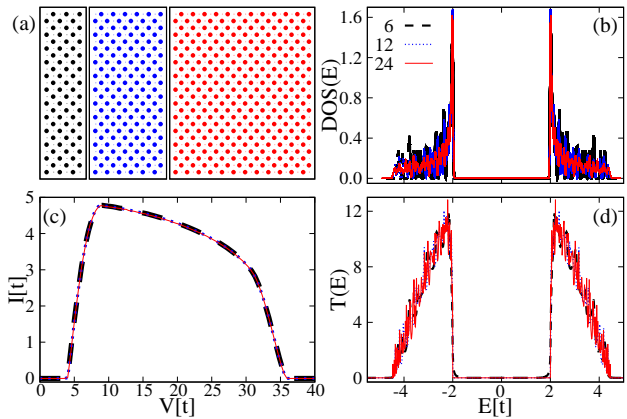}
    \caption{Ground-state $f$-particle configurations (a), DOS of the central system calculated with artificial Gaussian broadening with $\sigma=0.1$ (b),
    $I-V$ characteristics for systems (c) and equilibrium transmission function (d) at PHS point ($n_f=n_d=0.5$) and for system widths $L_x=6$ (black circles and lines), $12$ (blue), and $24$ (red) and $L_y=24$.
    \label{fig:AllConfig12}}
\end{figure}

As illustrated in Fig.~\ref{fig:AllConfig12}(a), 
the ground state at $\overline{\mu}=U/2$ is a perfect checkerboard ordering of $f$ particles for any system thickness. 
We emphasize again that this ordering will not change after we couple the system to the leads. The system DOS, shown in Fig.~\ref{fig:AllConfig12}(b), contains a CDW gap of width $\Delta_{\text{PHS}}=U$ centered around
the Fermi level. 
This gap is reflected in the equilibrium transmission function in Fig.~\ref{fig:AllConfig12}(d).

As a consequence, and despite the broadening coming from the leads, 
the current in Fig.~\ref{fig:AllConfig12}(c) is negligible for small voltages $V < \Delta_{\text{PHS}}$,
even for $L_x=6$. 
A sharp increase of the current can be seen when $V>U$ because then the 
chemical potentials of the leads reach the edges of the gap and the voltage 
starts to probe the main bands of the transmission function. 
It is important to note that here as well as in all other cases the shape
of the transmission function depends only weakly on the voltage if $V<B/2$ \cite{Zonda2019}. 
Therefore, we can in most cases illustrate its main qualitative features 
and its influence on the current by showing the transmission function profile for $V=0$.

As a result of the finite bands of the leads, the $I-V$ characteristic 
is nonmonotonic. 
It drops when the overlaps of the occupied states in the left lead, the unoccupied states in the right lead, 
and the DOS of the system decrease.
This is a common feature to all $f$-particle concentrations and we are 
therefore not addressing it in detail for other cases.

Because of the perfect checkerboard ordering, the $I-V$ characteristic is practically 
independent of the layer thickness for $L_x \gtrsim 6$. 
The reason is that for a wide enough system, the checkerboard structure
of the $f$ particles is seen by the electrons as a perfect periodic potential without any additional scatterers.

Note that any disruption of the checkerboard ordering, which cannot be avoided 
when leaving the PHS point, leads to the formation 
of subgap states or bands in the system's DOS \cite{MaskaPRB2006,Tran2006,ZondaSSC2009, Zonda2019b}.
However, the transmission through these subgap states in the vicinity of the PHS point 
is negligible for a sufficiently broad system \cite{Zonda2019b}.
The main reason is the effect of $d$-electron localization \cite{Antipov2016} 
for states within the gap.

We next move away from this special case starting with two typical half-filled cases ($n_f+n_f\simeq 1$).
 \subsubsection{Half filling $n_f+n_d\simeq 1$ \label{HFT0}}

\begin{figure}[!h]
	\centering
	\includegraphics[width=0.95\linewidth]{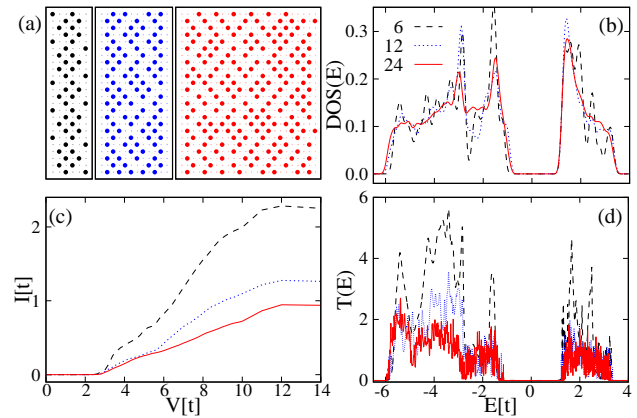}
	\caption{Groundstate charge configurations for $L_x=6$ (black circles), $12$ (blue), and $24$ (red) and $L_y=24$ (a), the equilibrium DOS of the decoupled system (b), current-voltage characteristics (c), and transmission function (d) for $n_f=1/3$, $\overline{\mu}=2.8t$ ($n_d\simeq 2/3$) at zero temperature. The color coding of the system width is the same for all panels.}
	\label{fig:AllConfig13hf}
\end{figure}

With lowering the concentration of $f$ particles to $n_f=1/3$ and fixing the equilibrium 
half-filling condition by $\overline{\mu}=2.8t$, 
the checkerboard phase starts to compete with the irregular diagonal stripes of unoccupied lattice points [Fig.~\ref{fig:AllConfig13hf}(a)]. 
The system has a tendency to keep 
the interfaces unoccupied or only sparsely occupied by the $f$ particles. 
This tendency seems to be rather general for the half-filled cases away from
PHS point. We have observed it also for higher concentrations of $f$ particles, which we 
do not discuss here for the sake of brevity.  

The disruption of the checkerboard pattern by the diagonal structures 
leads to formation of a sub-$\Delta_{\text{PHS}}$ band ($-2.5t\gtrsim E \gtrsim -1t$ in Fig.~\ref{fig:AllConfig13hf}(b,d)), 
which merges with the lower main band forming a strongly 
asymmetric DOS with a narrower gap $\Delta\sim U/2 \sim \Delta_{\text{PHS}}/2$. 
This is reflected in the $I-V$ characteristics, where the current starts to increase significantly
at much lower voltages than for the PHS case in Fig.~\ref{fig:AllConfig12}(c) and has a flatter slope for $2t<V\leq 10t$. 
Nevertheless, because of the disruption of the checkerboard pattern, the current in this range of voltages 
is smaller than in the PHS case, and it decreases with increasing system thickness for any $V$.

\begin{figure}[!h]
	\centering
	\includegraphics[width=0.95\linewidth]{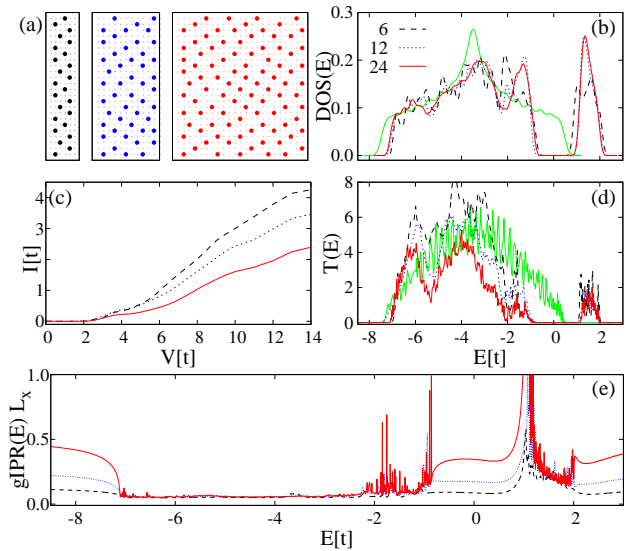}
	\caption{Ground state charge configurations for $L_x=6$ (black circles), $12$ (blue), and $24$ (red) and $L_y=24$ (a), the equilibrium DOS of the decoupled system (b), current-voltage-characteristics (c) and transmission function (d) for $n_f=1/6$, $\overline{\mu}=3.5t$ ($n_d\simeq 5/6$) at zero temperature.
    For comparison, the DOS (b) and transmission function (c) for an empty lattice at the same $\bar{\mu}$ are plotted by green line. The gIPR dependence on $E$ illustrating the different character of the states at low and high energies is plotted in (e). The color coding of the system width is the same for all panels.
	\label{fig:AllConfig16hf}}
\end{figure}

At very low concentrations of $f$ particles, represented in Fig.~\ref{fig:AllConfig16hf}
by the case $n_f=1/6$ and $\overline{\mu}=3.5t$, the ground state is a homogeneous distribution of
the localized $f$ particles. The gap around the Fermi level has even smaller width than for $n_f=1/3$ but it is still present.

Interestingly,  
the states in the vicinity of this gap have a relatively low transmission function 
when compared to the ones for $E<-3t$  (see Fig.~\ref{fig:AllConfig16hf}(d)), even though the system DOSs have similar magnitudes. 
This points to a different nature of the states in these two energy regimes,
at least for finite systems.
A similar difference can also be observed for the case $n=1/3$, 
but here it is much more pronounced, and therefore more suitable for a closer examination.  

The DOS for low energies $E<-3t$
resembles the noninteracting case.
We illustrate this in Fig.~\ref{fig:AllConfig16hf}(b) where we compare the DOS of the interacting case with that of a noninteracting system for $L=24\times 24$ (green line). 
It suggests that the states with low energies are less affected by the $f$ particles and therefore less localized. 
To support this conjecture, we plot in Fig.~\ref{fig:AllConfig16hf}(e) the gIPR$(E,w)$ scaled by $L_x$ for the three lattice widths used. 
Clearly, the scaling in the range of $-7t<E<-3t$ points to delocalized or only very weakly localized states. 
Unfortunately, the gIPR around the gap has too many sharp features to draw a reliable conclusion about 
how it evolves with increasing system size, 
but for a finite width it clearly indicates a higher localization than for the states in the range $-7t<E<-3t$.
The presence of a broad range of fairly delocalized states is the main reason why the current for $V>6t$  
shown in Fig.~\ref{fig:AllConfig16hf}(c) exceeds the one in Fig.~\ref{fig:AllConfig13hf}(c) for all
comparable lattice sizes, but most profoundly for the widest one.

Overall, the following conclusions can be drawn for the half-filled case. 
The open boundary conditions 
lead to $f$-particle configurations with the tendency to 
form unoccupied $f$-particle structures at the interfaces and 
the configurations have the same character for all lattice sizes. 
All cases studied, including those not presented here, exhibit a gap around the Fermi level in the system DOS
and have, therefore, a typical insulating character with negligible current for small voltages. 
The formation of bands within the former CDW gap can lower the threshold voltage at which a significant charge current starts to flow 
through the system but 
a localization of $d$ electrons seems to play a role in suppressing
the transmission for these states. 
\subsubsection{Away from half filling $n_f\simeq n_d$ \label{sec:ncT0}}

\begin{figure}[!h]
	\centering
	\includegraphics[width=0.95\linewidth]{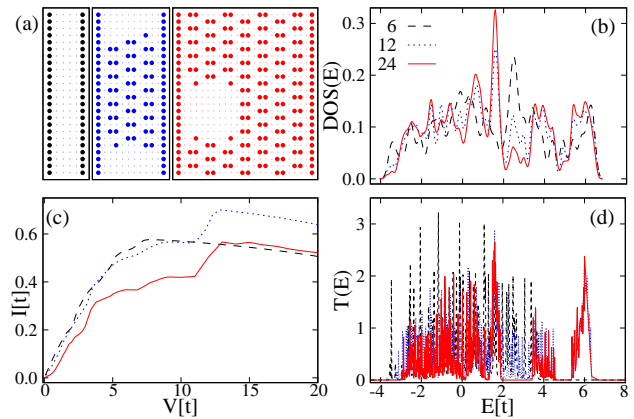}
	\caption{Ground state charge configurations for $L_x=6$ (black circles), $12$ (blue), $24$ (red) and $L_y=24$ (a), the equilibrium DOS of the decoupled system (b), current-voltage-characteristics (c), and transmission function (d) all for $n_f=1/3$ and $\overline{\mu}=-0.15t$ ($n_d\simeq 1/3$) at zero temperature. The color coding of the system width is the same for all panels.}
	\label{fig:AllConfig13n}
\end{figure}

A lowering of the total concentration of the particles 
below half filling leads to
various qualitative changes in the properties of the heterostructure.
First, the size of the system starts to  influence the 
character of the $f$-particle configurations significantly.
This can be seen in Fig.~\ref{fig:AllConfig13n} where we show data for $n_f=1/3$ and $\overline{\mu}=-0.15t$. These parameters set the neutral case $n_f\simeq n_d$.

For $L_x=6$ (and any smaller width), the $f$ particles are forming wall-like stripes  
at the edges of the system. The reason is the open boundary conditions at the interfaces for the 
decoupled system, which lead to a smaller mobility of the $d$ electrons 
at the interfaces than in the central part. 
Because of the low density, $n_d \sim 1/3$, of $d$ electrons it is energetically advantageous 
to push the $f$ particles to the edges and let the $d$ electrons concentrate in the empty 
central part of the lattice. 
This is, therefore, an example of a segregated phase.

Naturally, the walls of $f$ 
particles at the system-lead interfaces have consequences for the charge transport of the coupled system.
Because of the Coulomb interaction with the conducting electrons, the stripes
form a barrier which lowers the total magnitude of the current (compare Fig.~\ref{fig:AllConfig13n}(c) with  Fig.~\ref{fig:AllConfig13hf}(c)). 
However, there is also a benefit. 
The center of the lattice is empty which leads to a gapless DOS. 
The current has a metallic-like character with 
approximately linear dependence of the current on voltage for $V\lesssim 6$. 
The states above $E=3.5t$ in DOS have a negligible transmission for $L_x=6$ [the black line in Fig.~\ref{fig:AllConfig13n}(d) is practically at $T(E)=0$ in this range] and, therefore,  
do not contribute to the transport. Consequently, the $I-V$ characteristic
has its maximum already at $V\sim7t$ and the current declines way before 
it reaches the edges of the system DOS.

As the width of the system is increasing, the character of the configuration changes. 
The edges stay occupied, but a regular pattern of
$f$-particle dimers is formed in the central part of the system. 
The dimers are not spread in a most homogeneous way but concentrate in a domain with the rest of the lattice staying empty. 
This is therefore an example of a phase separation with three domains, namely,
fully occupied edges, the empty domain, and the dimer domain. 
This pattern does not change with further broadening of the system;
the dimer domain just becomes more dominant. 

The dimer domain does not open a gap in the DOS at the Fermi level; therefore,
the $I-V$ characteristic is still metallic-like,
even for $L_x=24$ (Fig.~\ref{fig:AllConfig13n}(c)). 
There are two interesting conclusions concerning the influence of the 
dimer domain on charge transport. 
First, the dependence of the current on the system width is rather weak 
for small voltages. This means, that the regular dimer pattern
is not a significant source of scattering for low energies.  
Second, this domain boosts the transmission   
for high energies $E>5t$ (blue and red lines in Fig.~\ref{fig:AllConfig13n}(d)). 
A clearly separated band  with high transmission is formed in the range $5t\lesssim E\lesssim 6.5t$
for systems that contain the dimer domain.
This band leads to a step-like increase of the current at $V\sim 12t$ and, surprisingly, the currents 
for broader system widths can therefore overcome even the one for $L_x=6$.

The reason for the high transmission through high energy states seems to be
that if the dimer domain spreads from one edge
of the system to the other, it is a perfectly periodic pattern (within the system).
This allows electrons to travel through system without scattering. 
Therefore, the dimer domain actually boost the current and the
only real barriers are the walls at the edges of the system, which are not
part of the dimer periodic pattern.

\begin{figure}[!h]
    \centering
    \includegraphics[width=0.7\linewidth]{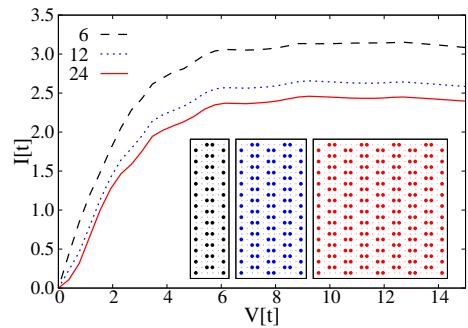}
    \caption{Current-voltage characteristics for the artificially ordered central systems shown in the insets with the lattice width $L_x=6$ (black circles), $12$ (blue), and $24$ (red) and $L_y=24$. All parameters besides the $f$-particle orderings are identical to the case shown in Fig.~\ref{fig:AllConfig13n}.}
    \label{fig:ArtDM}
\end{figure}

To show the magnitude of the influence of these walls on the transport, we plot
in Fig.~\ref{fig:ArtDM} the $I-V$ characteristics for systems with the
same $n_f$ and $\overline\mu$, where instead of the edge stripes a perfectly 
ordered pattern of dimers is formed. 
Note that although these orderings are artificial 
for open boundary conditions, they have been observed for periodic ones \cite{Zonda2012}
and might represent a configuration which stabilizes when using other measurement 
protocols, e.g., when the $f$-particle configuration orders in an already coupled system.  
The maxima of the current in Fig.~\ref{fig:ArtDM} are approximately 
five times higher than in Fig.~\ref{fig:AllConfig13n}(c), so we can conclude
that the effect of the edge walls is significant. 

As we lower the particle concentration even further, 
the segregation starts to dominate with
$f$ particles forming walls at the system edges for any width.
Because this leads to a relatively simple $I-V$ characteristic we omit 
this case from the discussion and instead  
\begin{figure}
    \centering
    \includegraphics[width=0.95\linewidth]{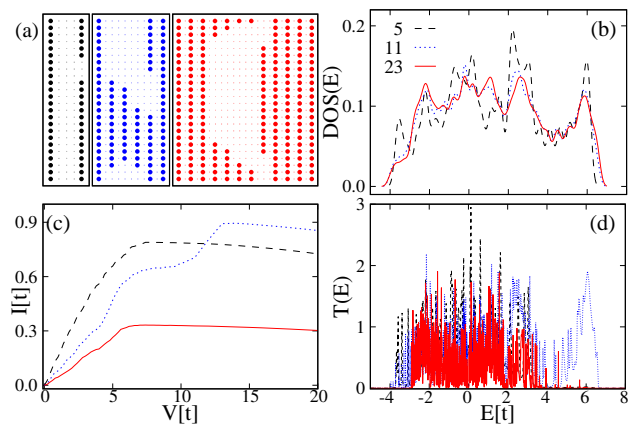}
    \caption{Ground state charge configurations for $L_x=5$ (black circles), $11$ (blue), $23$ (red) and $L_y=24$ (a), the equilibrium DOS of the decoupled system (b), current-voltage-characteristics (c) and transmission function (d) all calculated for $n_f=3/8$ and $\overline{\mu}=0.14t$ ($n_d \simeq 0.34$) at zero temperature. The color coding of the system width is the same for all panels.}
    \label{fig:AllConfig38n}
\end{figure}  
we address transport 
through a predominately charge-stripe ordered phase.

To this end, we investigate a ground-state configuration
where every other vertical layer is empty. Because both system edges tend to be 
occupied by the $f$ particles this type of ordering forms more naturally in 
systems with odd $L_x$, especially for narrow ones.
Inspired by Refs.~\cite{Lemanski2004,Zonda2012}, we choose $n_f=3/8$ and 
$\overline{\mu}=0.14t$ which stabilizes $n_d\sim 0.34$. 
As for $n_f=1/3$, the
smallest system shows a segregated phase, but from $L_x=11$ up clear vertical charge stripes
are formed, as shown in Fig.~\ref{fig:AllConfig38n}(a). These share the lattice with connected empty domains.

The $I-V$ characteristics in this case are metallic-like. 
The current through the system with $L_x=5$ and $L_x=11$ resembles the situation for $n_f=1/3$,
where $I$ for $L_x=12$ was at high voltages higher than $I$ for $L_x=6$.
However, the $I-V$ characteristic for $L_x=23$ differs qualitatively from the 
$L_x=11$ case. 
This can be seen already in the transmission function, Fig.~\ref{fig:AllConfig38n}(d), 
where the transmission function for $L_x=23$ (in contrast to $L_x=11$) 
does not show a clear band around $E=6t$.
The reason is that the upper band in the transmission for $L_x=11$ 
is related to the perfect stripe order in the lower half of the system, shown in the second panel of Fig.~\ref{fig:AllConfig38n}(a). Hence, it is related to the 
areas (channels) of perfectly periodically 
arranged $f$ particles spreading from the left to right interface,
where here, and in contrast to the dimer domain in Fig.~\ref{fig:AllConfig13n}(a), 
even the stripes at the edges play along. 

\begin{figure}[!h]
    \centering
    \includegraphics[width=1.0\linewidth]{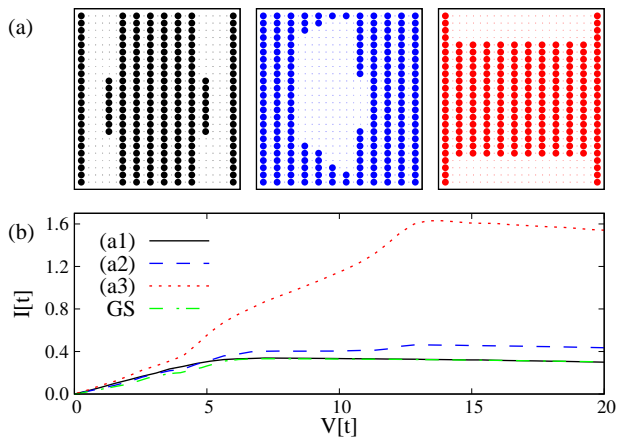}
    \caption{Artificial stripe configurations a1, a2, and a3 (from left to right) (a) at filling $n_f = 3/8$, $\bar{\mu}=0.14$, $L=23\times 24$, and their corresponding $I-V$ characteristics (b) illustrating the influence of complete periodic horizontal lines on the transport. The green dot-dashed line represents the ground-state solution. The color coding in panels (a) and (b) is the same.}
    \label{fig:Art38}
\end{figure}

Because an equivalent channel is missing for the system $L_x=23$, both the 
transmission at high energies and the current at high voltage 
drop significantly for this system width. 
To test this conjecture we show in Fig.~\ref{fig:Art38} the $I-V$ 
characteristics of some artificial stripe orderings 
for the same $n_f$ and $\overline{\mu}$.
Panel (a1) represents a case without a periodic channel. 
Its $I-V$ characteristic at high voltages matches the current through the true ground-state ordering (plotted by the green dashed line for comparison). The case shown in (a2) represents a slight modification of the ground-state
configuration which now has three perfectly periodic (within the system) horizontal lines. 
Its $I-V$ characteristic already contains the step-like increase of the current at high voltages. 
This also shows that a small $f$-particle excitation from the ground state, can have
a big influence on the current at high voltages.  
Finally, panel (a3) represents an optimal case
with keeping the completely occupied edges. Its current at high voltages is more than four times higher than for the
the ground state, showing again the importance of the periodic arrangements. 

In general, the cases with low total concentration of particles are more sensitive to the system width and the open boundaries than
the half-filled one. 
Because of that, the $f$-particle orderings and the $I-V$
characteristics have different 
character for small and large system widths. 
Although the $I-V$ characteristics are metallic-like for all investigated system widths, 
the periodic channels connecting left and right interfaces, observed only for wider systems, can significantly enhance the current at high voltages. 

\subsection{Finite temperatures \label{FinT}}
The finite-temperature phase diagram of the two-dimensional FKM is surprisingly rich 
even at the PHS point \cite{Antipov2016}. 
It describes, in the thermodynamic limit and at high temperatures, either a Mott insulator ($U\gtrsim 7t$), characterized by a finite gap in the DOS, 
or a gapless Anderson insulator ($U\lesssim 7t$). However,
for finite systems the latter phase actually consists of a smooth
crossover from the Anderson insulator ($U\lesssim 7t$) through a weakly localized phase, 
which can have a metallic-like character, down to a Fermi gas at $U=0$ \cite{Antipov2016}. 
Below a critical temperature $\tau_c$,
which depends on $U$ and has a maximum of $\tau_c\approx 0.16t$ at $U\approx 3.5t$, the system enters an ordered phase. This phase shows a broad CDW gap between the main Hubbard bands 
in the DOS.  
Nevertheless, at finite temperatures this gap also contains subgap states 
emerging as a consequence of the $f$-particle excitations \cite{MaskaPRB2006,Zonda2019b}.
\begin{figure}[!h]
	\centering
	\includegraphics[width=1.0\linewidth]{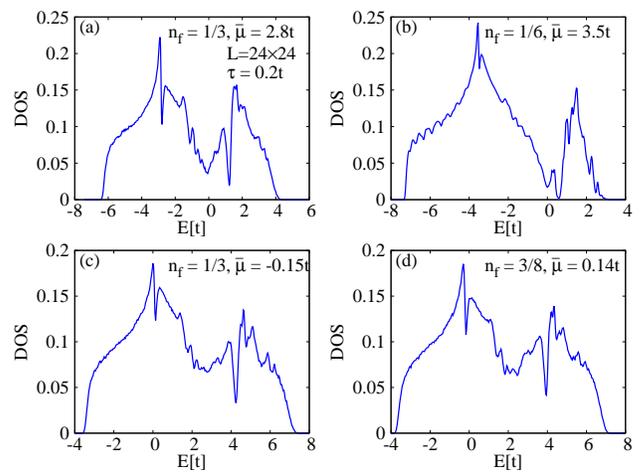}
	\caption{System DOS at high temperature ($\tau=0.2t$) for $n_f=1/3$ $\overline\mu=2.8t$ ($n_d\simeq 2/3$) (a), $n_f=1/6$ $\overline\mu=3.5t$ ($n_d\simeq 5/6$) (b),  
	$n_f=1/3$ $\overline\mu=-0.15t$ ($n_d\simeq 1/3$) (c), and $n_f=3/8$, $\overline\mu=0.14t$ ($n_d\simeq 0.34$) (d). }
	\label{fig:DOSht}
\end{figure} 

The situation outside the PHS point is even more complicated \cite{Lemanski1995,LemanskiPRL2002,Lemanski2004,Tran2006,Zonda2012}. 
As illustrated in Fig.~\ref{fig:DOSht}, the high-temperature DOS for $U=4t$, on which we focus, is gapless for all investigated fillings. 
Therefore, we can exclude the Mott insulator phase from our analysis.
 
With decreasing temperature the system often goes
from the disordered phase first
through some transient orderings before the critical temperature of the low-temperature ordered phase is reached \cite{Tran2006,Zonda2012,Zonda2019}.
In the vicinity of the PHS this transient phase usually shows CDW orderings \cite{Tran2006,Zonda2012}. 
Far away from the PHS the transient orderings typically 
form for systems where the ground state is phase separated. 
This happens because some of the constituent phases can start to order at higher temperatures than others. These transitions and crossovers have a significant influence on the
nonequilibrium charge transport.

We illustrate this by focusing on three representative 
cases. The first, with $n_f=1/3$ and $\overline{\mu}=2.8t$ (half-filled case), is insulating at zero temperature and shows a ground-state ordering where a checkerboard structure competes with diagonal stripes. The remaining two cases, $n_f=1/3$, $\overline{\mu}=-0.15t$ and $n_f=3/8$, $\overline{\mu}=0.14t$, are metallic but with fully occupied interfaces and different $f$-particle orderings in the central part.

\subsubsection{Insulating case}
\begin{figure}[!ht]
	\centering
	\includegraphics[width=1.0\linewidth]{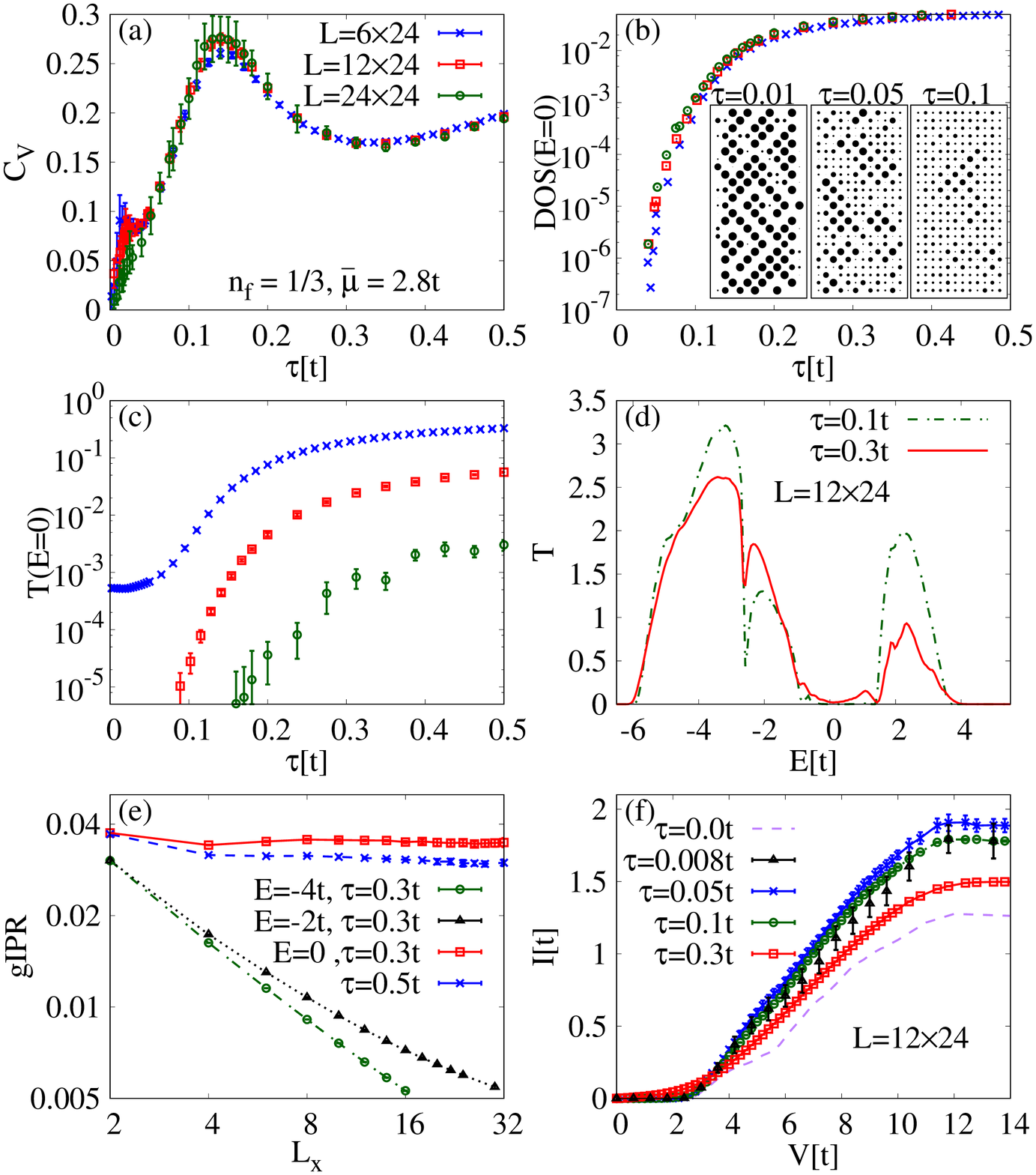}
	\caption{The half-filled case $n_f=1/3$ fixed by $\overline{\mu}=2.8t$. Figures show the specific heat as a function of temperature (a), averaged $f$-particle densities (see the text) (b), the transmission 
	function at Fermi level as a function of temperature (c), the transmission function (d), gIPR for chosen energies at high temperatures (note that both scales are logarithmic) (e), and $I-V$ characteristics for $L=12\times 24$ and various temperatures (f).}
	\label{fig:FT13hf}
\end{figure}

The half-filled case, $n_f=1/3$ and $\overline{\mu}=2.8t$, is 
an example of a system with a transient regime. 
This can be seen in Fig.~\ref{fig:FT13hf}(a) where we plot
the specific heat, $C_V$, as a function of temperature. 
The specific heat shows 
two low-temperature anomalies (local maxima) 
for all three system widths addressed.
Together with the three examples of averaged configurations in the inset of Fig.~\ref{fig:FT13hf}(b), these anomalies point to two stages of the ordering process.
The checkerboard domains start to form already below $\tau\approx 0.12t$, and only 
at much lower temperatures ($\tau\approx 0.01t$)
the irregular axial empty stripes and 
the empty structures at the vicinity of the interfaces, found in the ground-state configurations in Fig.~\ref{fig:AllConfig13hf}(a), start to form.
The $f$-particle densities in the inset of Fig.~\ref{fig:FT13hf}(b) were calculated from a 
single MC run and averaged over hundred successive measurements separated by a single sweep. 
Therefore, they can be seen as thermally blurred snapshots of
a typical configuration.

Note that we use the anomalies in $C_V$ together with some supporting quantities, like typical configurations or the order parameters, only to 
estimate the temperatures below which an 
ordering starts to form for a specific system width and  
to see how the orderings depend on the temperature. 
We do not investigate whether the anomalies relate in the thermodynamic limit to 
crossovers or real phase transition. 
For that we would have to resort to a finite-size scaling for much larger systems than we can currently access. 

We start the analysis of the transport properties in the disordered phase 
at high temperatures. Here, the system has a
high DOS at the Fermi level which does not depend on the system width (see Fig.~\ref{fig:DOSht}(a) and Fig.~\ref{fig:FT13hf}(b)). 
However, the transmission function at the Fermi level rapidly decreases with the system width. This 
can be seen in Fig.~\ref{fig:FT13hf}(c) where $T(E=0)$ at $\tau=0.5t$ is for 
system width $L_x=24$ more than a hundred times smaller than for $L_x=6$. In addition,
a clear pseudo gap can be seen in the transmission function 
(Fig.~\ref{fig:FT13hf}(d)) 
in the vicinity of the system's Fermi level.
These qualitative differences in DOS and $T$ point to a relatively strong localization of the states in the vicinity of the Fermi level.
This is also supported by the gIPR in Fig.~\ref{fig:FT13hf}(e), which at
$E=0$ quickly saturates and stays practically constant with increasing $L_x$ for both $\tau=0.3t$ and $0.5t$. 
Such a saturation is typical for localized states and is in contrast to the
situation for $E=-2t$ and $E=-4t$, where the gIPR at $\tau=0.3t$ 
does not show any sign of localization for 
the addressed system widths (note the logarithmic scales).

The formation of the checkerboard domains, which takes place below $\tau\sim 0.12t$, leads to a 
sharp decrease of the DOS at the Fermi level (Fig.~\ref{fig:FT13hf}(b)).
This is accompanied by a step-like [note the logarithmic scale in Fig.~\ref{fig:FT13hf}(c)] decrease
of the 
transmission function at the Fermi level
and with the broadening of the deep pseudogap in the transmission function in Fig.~\ref{fig:FT13hf}(d). 
The transmission function actually
decreases in the range of energies from $\sim-2.5t$ to
$\sim 1.5t$ which reflects the CDW gap $\Delta_{\text{PHS}}=U$ of the PHS case. 
On the other hand, the transmission through the main bands is increasing.
This results in different tendencies for the current at low and high voltages already observed for the PHS case \cite{Zonda2019}.
For $V<4t$ the current decreases with decreasing temperature, which reflects the formation of the gap in the DOS and transmission function. For $V>4t$, the current first increases, which follows the increasing transmission through the main bands and reflects the formation of periodic CDW structures. However, it decreases for the temperatures below the position of the lower anomaly in $C_V$ where the labyrinth-like patterns of 
empty diagonal stripes disrupt the checkerboard
configuration of $f$ particles.

The broad pseudo gap of width $\sim 2t$ dominates the $I-V$ characteristics at low temperatures and
small $V$. Because this gap forms already below $\tau=0.12t$, 
the transition to the phase with clear domains
which takes place below $\tau=0.01t$ does not lead to a qualitative difference in the current 
profile at small voltages. The transmission function $T(E=0)$ has the tendency to saturate to its 
zero-temperature value which is most evident for a small system width.
However, even for this case the transmission function is in the ordered phase more than three orders
of magnitude lower than in the disordered phase. 
This leads to a negligible current for small voltages even for $L_x=6$ (see Fig.~\ref{fig:AllConfig13hf}). As a result, the finite-temperature $I-V$ characteristics
resemble even at low temperature the ones for the PHS case \cite{Zonda2019}, although,
with smaller gap. 

\subsubsection{Metallic-like cases}
The cases  $n_f=1/3$, $\overline{\mu}=-0.15t$ and $n_f=3/8$, $\overline{\mu}=0.14t$ show 
metallic-like transport properties even
for zero temperature and broad system width. The main reason is that the 
Fermi level is shifted into the lower main band of the systems DOS, where the states are fairly delocalized.
At finite temperatures these cases also go through some intermediate 
orderings, which influence the transmission. 
\begin{figure}[!h]
	\centering
	\includegraphics[width=1.0\linewidth]{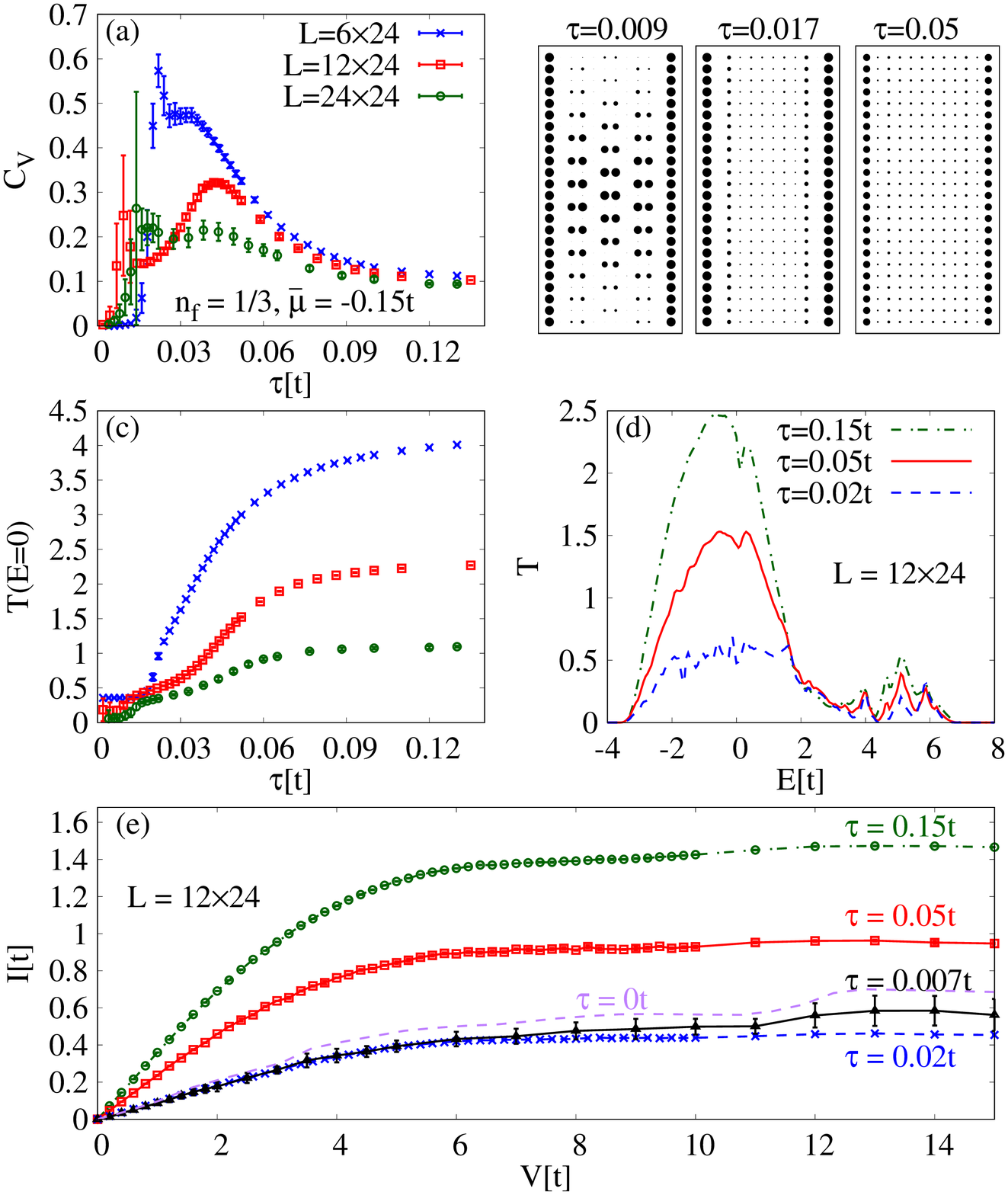}
	\caption{Thermodynamics and transport properties for the case $n_f=1/3$ and $\overline{\mu}=-0.15t$ ($n_d \simeq 1/3$). Figures show the specific heat as a function of temperature (a), averaged $f$-particle densities (b), the transmission 
	function at system Fermi level as a function of temperature (c), the transmission function for chosen temperatures (d), and $I-V$ characteristics for $L=11\times 24$ and various temperatures.}
	\label{fig:FT13n}
\end{figure}

Figure~\ref{fig:FT13n} shows the case  $n_f=1/3$ and $\overline{\mu}=-0.15t$.
For small system widths the ground state is a segregated phase 
with $f$ particles ordered at the interfaces. 
For wider system width an additional dimer domain exists in the central part. 
However, the ordering at the edges starts to form at significantly higher 
temperatures ($\tau\approx 0.045t$ for $L_x>6$) than the dimers ($\tau\approx 0.01t$). 
This can be seen in the evolution of the thermally blurred snapshots in Fig.~\ref{fig:FT13n}(b). 
The formation of the isolated stripes at the edges (combination of fully occupied interface and empty next layer) 
leads to a broad local maximum in specific heat that decreases with the system width.
Because this ordering represents an effective barrier for the conducting electrons,
its formation leads to a significant decrease of the transmission with decreasing temperature as shown in Fig.~\ref{fig:FT13n}(c,d). 

This has a paradoxical 
consequence for the $I-V$ characteristics shown in 
Fig.~\ref{fig:FT13n}(e). Although the $I-V$ characteristic has a metallic-like
character for all temperatures and there is no gap forming at the Fermi level, 
the current actually decreases with decreasing temperatures for all $V$. 
This changes only when the central part starts to order below
$\tau\approx 0.01t$ as signaled by low-temperature anomalies in $C_V$ for
$L_x=12$ and $24$.
We illustrate this by the $I-V$ characteristics for $\tau=0.007t$, which
follows the $\tau=0.02t$ curve at low voltages 
but exceeds it at high voltages approaching the
zero-temperature result (dashed purple line). The former shows that the formation on the edges is 
the major cause for the lowering of the transmission around the 
Fermi level.
The latter reflects the ability of the periodic channels to boost the 
transmission at high energies as discussed in Sec.~\ref{sec:ncT0}.

\begin{figure}[!h]
	\centering
	\includegraphics[width=1.0\linewidth]{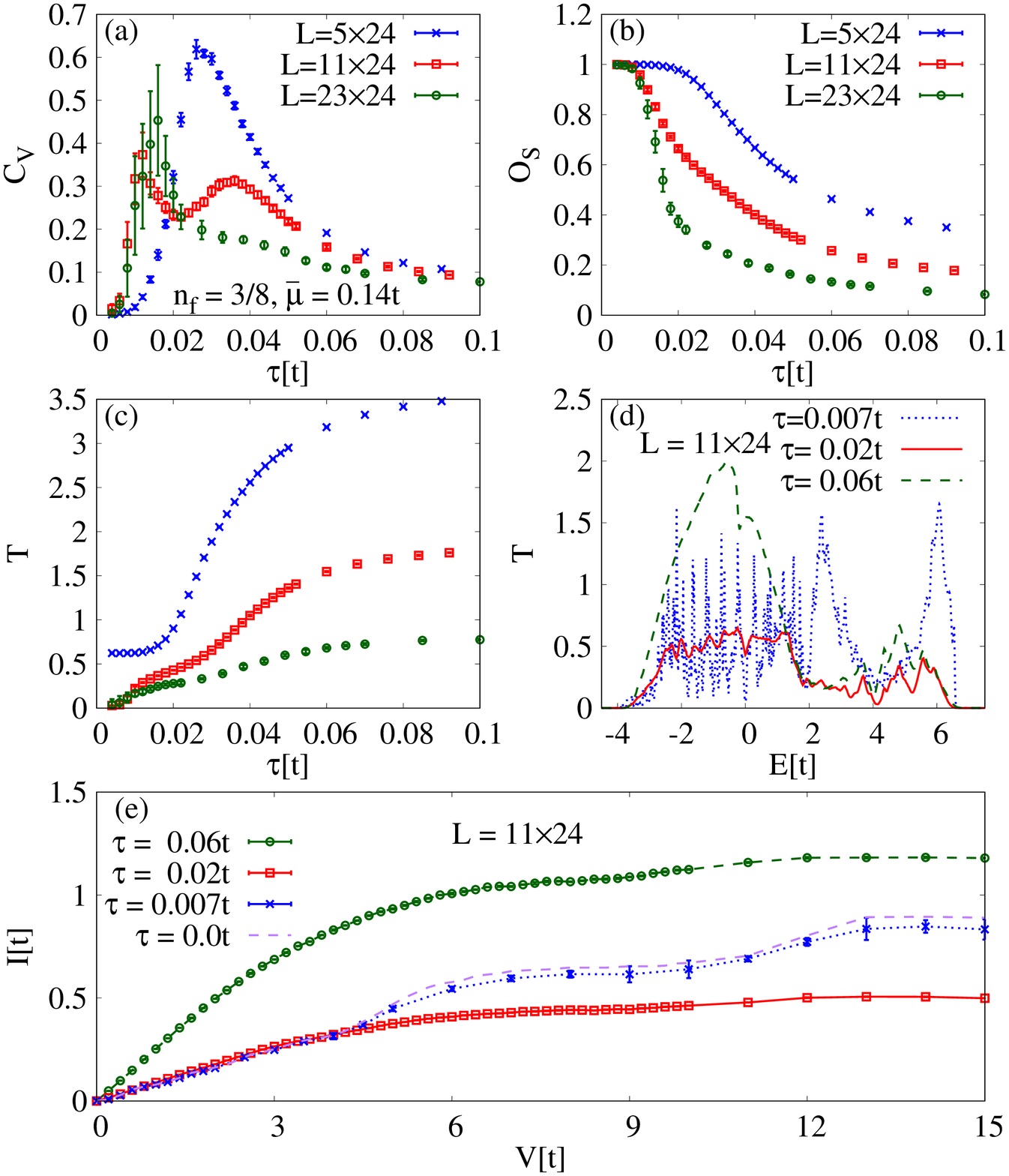}
	\caption{Results for $n_f=3/8$ and $\overline{\mu}=0.14t$. Figures show the specific heat as a function of temperature (a), 
	the order parameter for the vertical stripe phase defined in the text (b), the transmission 
	function at the system's Fermi level as a function of temperature (c), the transmission function (d) and $I-V$ characteristics for $L=12\times 24$ and various temperatures (e).}
	\label{fig:FT38n}
\end{figure}

The last case we investigate in Fig.~\ref{fig:FT38n} is characterized by $n_f=3/8$ and $\overline{\mu}=0.14t$. The filling is close to the previous 
case and they therefore share some properties, including the 
formation of the $f$-particle walls at the interfaces before the rest of the system starts to order. 
This can be seen as maxima in the $C_V$ (at $\tau\approx 0.03t$) which decrease with
decreasing system width and coincide with the decreasing transmission at zero energy shown in  Fig.~\ref{fig:FT38n}(c,d).

The regime where the $f$-particle wall is already formed can be
seen also in the changed slope of the $T(E=0)$ for $\tau<0.03t$.
At lower temperatures the vertical 
stripes start to form in the whole system if it is wide enough. 
This leads to formation of the clear maximum in $C_V$ at $\tau\approx 0.015t$
shown in  Fig.~\ref{fig:FT38n}(a), which correlates with a 
sharp increase of the
order parameter for stripes defined as 
\begin{equation}
    O_S=\frac{1}{N_f}\sum_{j}^L (-1)^{l_x(j)-1}w_j,
    \label{O_S}
\end{equation}
where $l_x$ numbers the layers from the left interface. 

The formation of the stripes enhances the transmission at 
the high-energy states for $L_x=11$ (blue dashed line in  Fig.~\ref{fig:FT38n}(d))
and boosts that way the current for high voltages.
It also leads to formation of two distinct band like structures
in the transmission function for $E>2t$ [Fig.~\ref{fig:FT38n}(d)],
which is reflected in the two step-like increases of 
the current, where already the first step overcomes 
the current calculated at more than two times higher temperature.
All this illustrates; that the formation of charge stripes can lead to a nontrivial charge transport characteristics in correlated electron systems.

In general, we conclude that the transient orderings can have a
different effect on the charge current. The half-filled case
illustrates that the formation of the checkerboard patterns leads
to the opening of a pseudo gap in the transmission function. 
As a consequence the current at small voltages rapidly vanishes 
with decreasing temperature; however, simultaneously the current at high 
voltages increases. The formation of the ground-state-like orderings does not
change the low-voltage tendency, but the high-voltage current
is decreasing.

The latter two cases differ from this picture. The formation 
of the $f$-particle walls at relatively high temperatures leads 
to suppression of the current with decreasing temperature for all voltages. 
This stops at temperatures where the center of the system starts to order and the formation of periodic structures can even lead to a significant increase
of the current at high voltages.

\section{Summary \label{Summary}}
We have studied the influence of nonhomogeneous charge orderings on the nonequilibrium  charge transport
in the
two-dimensional FKM  for a range of particle fillings and electrochemical potentials.
The main results of the study can be summarized as follows.

The ground-state configurations of the half-filled cases
are not very sensitive to the system width and 
show an insulating $I-V$ characteristics. 
Although, the transport properties of these configurations differ in details, 
qualitatively they resemble the PHS case.

A completely different situation was found for systems far away from half filling
where we focused on the neutral case.
Here, the transport was predominately metallic-like at any system width 
but the ground-state configurations, and consequently also $I-V$ characteristics, 
were sensitive to the system size.
Interestingly, periodic domains, e.g., vertical stripes of dimer domain,
which form only for sufficiently wide systems can significantly boost 
the transmission at high voltage if these domains spread from one system interface to the other. 
As a consequence, the current can actually increase
with the increasing system width. 

Addressing the finite-temperature 
properties for three representative cases,
we have shown that with decreasing
temperature the system tends to go 
from the disordered phase through various transient orderings before
it reaches the ground state.
These transient orderings significantly influence the transport properties
and lead to various complicated current-temperature dependencies.
For example, the formation of checkerboard domains suppresses the current at low voltages and enhance it at high voltages. 
On the other hand, the formation of $f$-particle walls 
suppresses the
current for all voltages with decreasing temperature, 
but this tendency is overruled by the formation of the periodic structures in the center of the system, which 
boost the current at high voltages.

Overall, the FKM away from the PHS point naturally shows a rich variety of stable charge orderings.
The typical cases investigated in the paper illustrate several qualitatively different, and sometimes nonintuitive, dependencies of the charge transport on voltage and temperature.
This shows the crucial influence of the nonhomogeneous charge orderings, such as stripes and dimer domains or phase separation, on the transport properties of correlated electron systems. 
It also illustrates the rich nonequilibrium physics of correlated electrons on a lattice, 
even when they are described by a simple model such as the spinless FKM. 
Qualitative changes can be obtained by varying the particle filling or potentials of the system.
This opens a vast variety of problems that can be addressed by the used method and various extensions 
of the FKM. Besides investigating other regimes of the spinless FKM, e.g., the vicinity of the PHS case
where various nonhomogeneous axial orderings compete with full CDW phase \cite{Zonda2012}, insight into the
spin-resolved current
can be gained from the investigation of the transport through the spinful version of the FKM, where 
the homogeneous charge orderings are accompanied by different spin orderings \cite{Lemanski2008,ZondaPhT2009,Wrzodak2010}. 
Such studies are expected to be particularly interesting in the context of spintronics. 

\begin{acknowledgments}
The authors acknowledge support by the state of Baden-W\"urttemberg
through bwHPC and the German Research Foundation (DFG) through grant
no INST 40/467-1 FUGG. 
\end{acknowledgments}

\section*{Appendix A - Stability of half-filled and neutral case}
\begin{figure}[!h]
    \centering
    \includegraphics[width=0.95\linewidth]{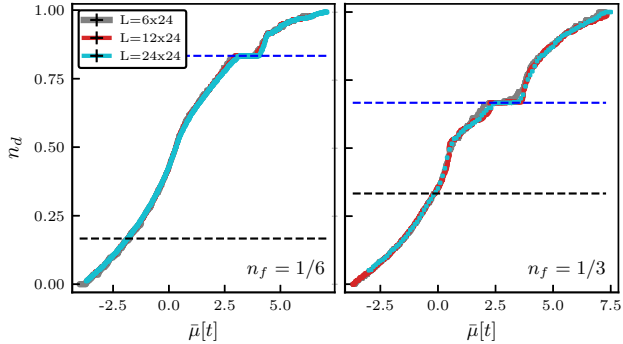}
    \caption{Ground-state dependencies of $d$-electron concentration on the electrochemical potential $\overline{\mu}$ for $n_f=1/6$ (left) and $n_f=1/3$ (right) and three lattice sizes. 
    Black dashed line depicts the neutral case $n_d=n_f$ and the blue dashed line the half-filled case $n_d+n_f=1$.}
    \label{fig:NdAll}
\end{figure}

In the first step of our investigation we had to find the electrochemical 
potentials which stabilize the half-filled and neutral cases of the isolated system. 
We used the simulated annealing to calculate the approximate 
$n_d -\overline{\mu}$ dependencies (for example see Fig.~\ref{fig:NdAll}),
then refined the calculation in the vicinity of the correct $n_d$.
By this procedure, we could approximate the 
ranges of $\overline{\mu}$ which stabilize the
wanted $n_d$.
In all cases, we found an 
overlap between ranges of $\overline{\mu}$ for different lattice sizes. Then 
we have chosen the value of $\overline{\mu}$ from this overlap 
and used it for all system sizes. 

\begin{figure}[!h]
    \centering
    \includegraphics[width=0.9\linewidth]{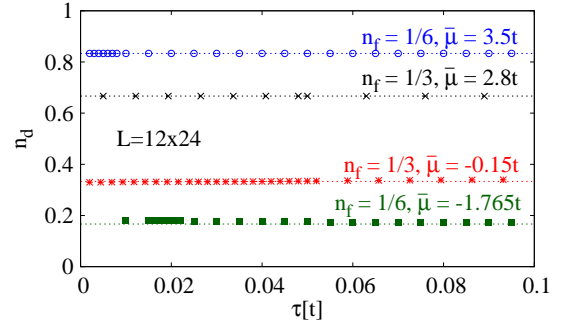}
    \caption{Itinerant-electron densities of the decoupled system plotted as a functions of temperature. The dotted lines represent the values $5/6$, $2/3$, $1/3$, and $1/6$.}
    \label{fig:ndtau}
\end{figure}

It is important to note, that the electrochemical potential strictly fixes the desired 
$n_d$ 
only at zero temperatures and zero voltage. Figure~\ref{fig:ndtau} shows,
how $n_d$ depends on the temperature for various fillings and $\overline{\mu}$ in the decoupled system. 
The $n_d$ for half-filled cases is perfectly stable with the increasing temperature 
(illustrated by  $n_f=1/6$ with $\overline{\mu}=3.5 t$ and $n_f=1/3$ with $\overline{\mu}=2.8 t$ in Fig.~\ref{fig:ndtau}).
To some extent, the same can be said about the neutral case, where for the presented cases and
studied temperatures the deviation from the $n_d$ at ground state was always below $1\%$.
\begin{figure}[!h]
    \centering
    \includegraphics[width=0.9\linewidth]{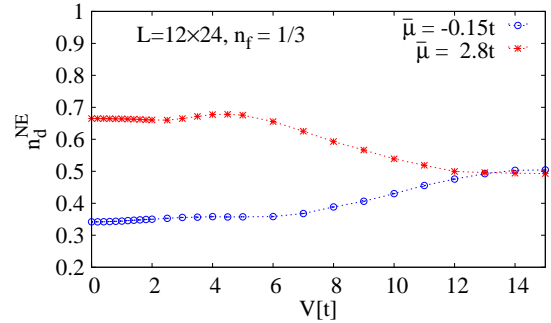}
    \caption{Nonequilibrium density of the itinerant electrons in the coupled system for $n_f=1/3$ $\overline{\mu}=-0.15t$  and $\overline{\mu}=2.8t$ calculated for $L=12\times 24$ and $\tau=0.1t$.}
    \label{fig:ndV}
\end{figure}

The situation becomes more complicated after we connect the leads and set a finite voltage. 
This is illustrated in Fig.~\ref{fig:ndV}, where we show the high temperature $n^{\text{NE}}_d$
as a function of $V$ for $n_f=1/3$, $\overline{\mu}=-0.15t$ and  $\overline{\mu}=2.8t$.
Both curves are approximately constant only for small voltages. 
The $n^{NE}_d$ for the $\overline{\mu}=-0.15t$ case starts to markedly increase at $V\gtrsim6t$ and the $n_d$ 
for the $\overline{\mu}=2.8t$ case
decreases at this range. This behavior can be qualitatively 
explained by focusing on the high-temperature system DOS showed in 
Fig.~\ref{fig:DOSht}(a,c). The DOS is quite asymmetric in both cases
and the crucial point here is the position of the $E=0$ in accordance to the distances from the left and right edges of the 
system DOS. 
For $\overline{\mu}=2.8t$, the zero is closer to the upper edge and there are fewer states above zero than below it. 
Therefore, as we increase the voltage; and thus shift the
DOS of the left lead up and right one down in energy, 
the $\mu_L$ reaches the upper edge at a much lower voltage ($V\sim 8t$) than 
at which $\mu_R$ reaches the bottom one ($V\sim 12$).
Consequently, because in this range of voltages there are no more states for the left 
lead to be occupied by electrons but still more states to be depleted from the right lead, 
the steady-state $n^{\text NE}_d$ decreases.
Because the case with $\overline{\mu}=-0.15t$ has $E=0$
much closer to the bottom than upper edge of the system
DOS it follows a reversed scenario from the $\overline{\mu}=2.8t$ case. 
\begin{figure}[!h]
	\centering
	\includegraphics[width=0.89\linewidth]{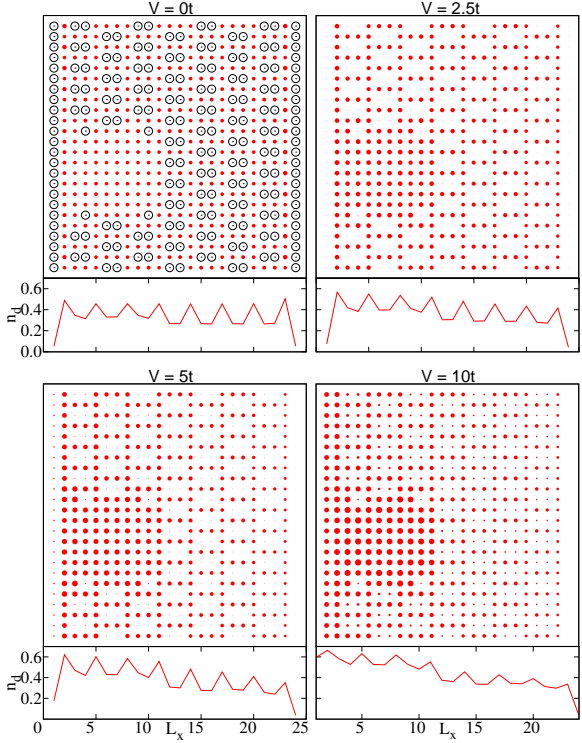}
	\caption{ Nonequilibrium distribution of the $d$-electron densities of the coupled system (red circles) 
	         for various voltages $L=24\times24$, $n_f=1/3$, and $\overline{\mu}=-0.15t$. 
	         The $V=0t$ case shows also the position of the $f$ particles (black circles), which does not depend on the voltage. 
	         The radius of the black circles represents occupancy $1$ and presents that way the scale for the red circles.}
	\label{fig:NE_D_dis13n}
\end{figure}

Because of the regular long range nonhomogeneous character of the $f$-particle configuration in Fig.~\ref{fig:AllConfig13n}(a) we decided
to use this case ($n_f=1/3$, $\overline{\mu}=-0.15t$) to illustrate the influence of the finite voltage on the nonequilibrium real-space distribution of the $d$ electrons. 
This is shown in Fig.~\ref{fig:NE_D_dis13n} where panel (a) represents the equilibrium case. 
Here, for comparison, we show also the position of the localized $f$ particles (black circles). 
It is obvious, that the interaction $U$ is pushing the $d$ electrons away from the sites occupied by $f$ particles. As we increase 
the voltage, and despite the fixed value of $\overline{\mu}$, an increasing left-right slope can be observed in the $d$-electron density calculated per layer (bottom panels). 
Nevertheless, the tendency of $d$ electrons to avoid the $f$ particles is clear for all voltages and shows that they also form inhomogeneous charge patterns.

\section*{Appendix B - Finite-size scaling in the $y$ direction}
\begin{figure}[!h]
    \centering
    \includegraphics[width=1.0\linewidth]{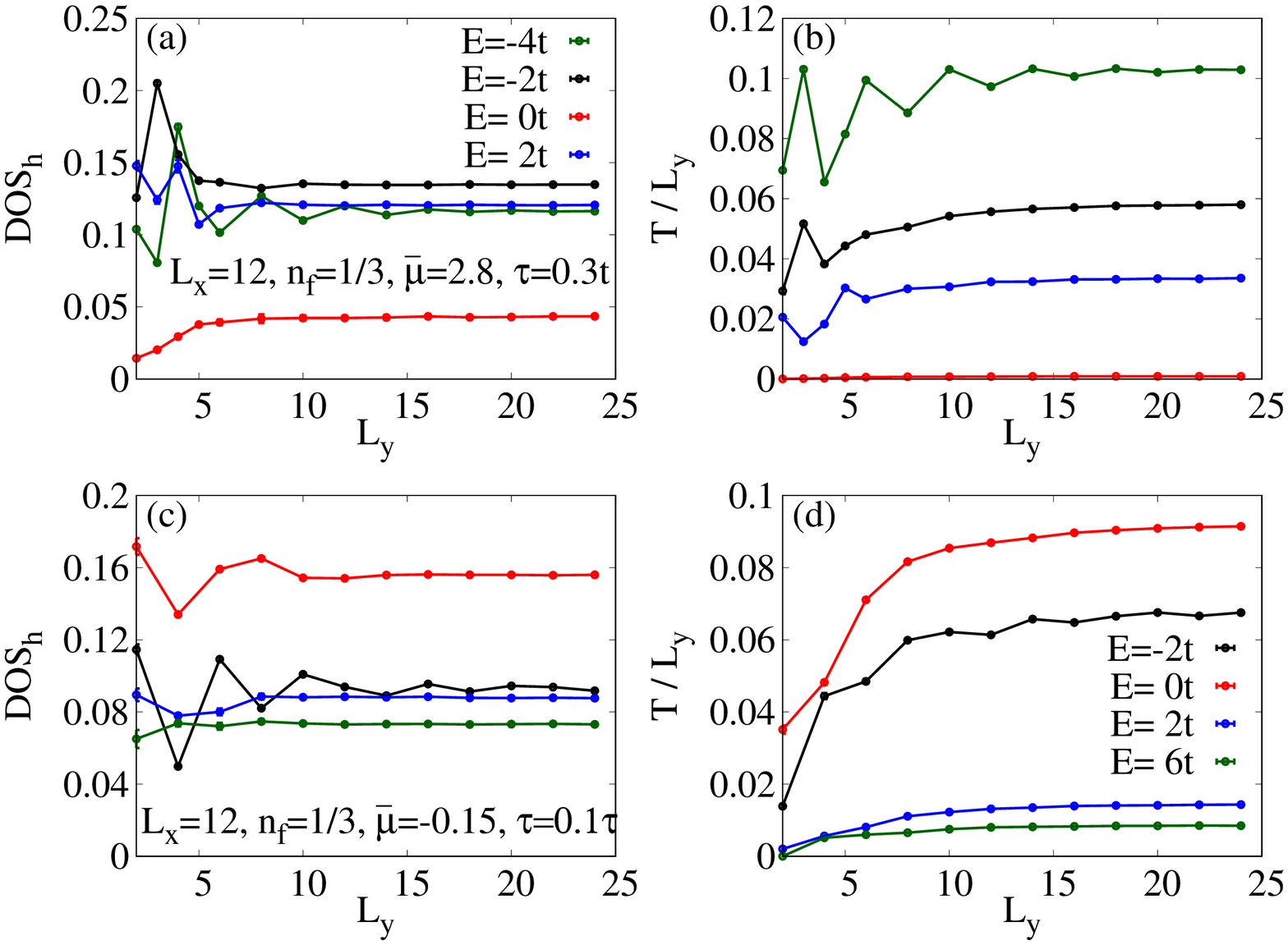}
    \caption{Finite-size scaling of the DOS [(a),(c)] of coupled system and scaled transmission function [(b),(d)] at various energies calculated for the case $n_f=1/3$ and $\overline{\mu}=2.8t$ at $\tau = 0.3t$ [(a),(b)] and 
the case $n_f=1/3$ and $\overline{\mu}=-0.15t$ at $\tau = 0.1t$ (c,d) for fixed $L_x=12$.}
    \label{fig:Ly}
\end{figure}
We restricted the vertical size of the system in all numerical calculations presented in the main text to $L_y=24$. This value was motivated by our recent studies focused on the transport in the PHS case \cite{Zonda2019,Zonda2019b}, where we showed that when 
using mixed boundary conditions a vertical size of 
$L_y\sim 20$ is sufficient for a reliable approximation of the 
$L_y\rightarrow \infty$ limit. A similar statement can be claimed also outside the PHS point. 
We show this in Fig.~\ref{fig:Ly} where we plot some examples
of the finite-size scaling on the $L_y$ for fixed $L_x=12$.
The left panels depict DOS$_h$ of the coupled system and the right ones the transmission function density calculated at various energies chosen from different bands.
The first example was calculated for $n_f=1/3$ and $\overline{\mu}=2.8t$ 
(half filled) at $\tau = 0.3t$  and the second for
 $n_f=1/3$ and $\overline{\mu}=-0.15t$ 
(neutral) at $\tau = 0.1t$.
Both examples show that the DOS$_h$ as well as $T$ 
rapidly converge with increasing $L_y$ and are practically saturated at $L_y=24$.

\section*{Appendix C - Influence of $f$ particle excitations on transport}

We have shown in Fig.~\ref{fig:FT13hf} that in some cases already a very small temperature ($\tau < 0.01t$) can
significantly modify the ground-state charge current at high voltages. The main reason is the excitation of the 
$f$ particles. Interestingly, low-temperature excitations can both  enhance  (Fig.~\ref{fig:FT13hf}) as well
as  suppress (Figs.~\ref{fig:FT13n},\ref{fig:FT38n}) the current at high voltages.
Although other processes are involved as well, e.g., localization, we have discussed in the main text that the major reason for the reduction of the current is the disruption of periodic patterns.
On the other hand, we have also shown that excitations which provide complete periodic channels to the $f$-particle configurations can lead to the enhancement of the current (see the discussion to Fig.~\ref{fig:Art38}).
\begin{figure}[!h]
    \centering
    \includegraphics[width=1.0\linewidth]{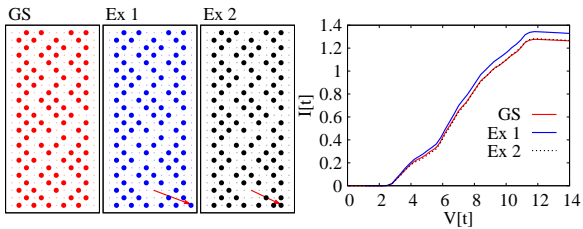}
    \caption{Comparison of $I-V$ characteristics of the system in the ground state (GS) ordering and GS configuration modified by different single-particle excitations (Ex~1,Ex~2). The excitations are marked by arrows. The transport was calculated for $n_f=1/3$, $\overline{\mu}=2.8t$ ($n_d\approx 2/3$), $U=4t$, and $L=12\times 24$.}
    \label{fig:Ex}
\end{figure}
Here, we readdress the problem for small but finite temperatures focusing on the excitations that are responsible for the enhancement of the current at low temperatures shown in Fig.~\ref{fig:FT13hf}(f).
 
Figure~\ref{fig:Ex} illustrates how moving a single $f$ particle can significantly enhance the current if the moved particle (indicated by an arrow in panel Ex~1) completes a periodic chain.  Note that with increasing temperature such excitations are
preferred because above $\tau\approx 0.01t$ the system orders in CDW domains.    

However, most of the possible single particle excitations actually have a negligible effect on transport (see, for example, the black dashed line in the rightmost panel of Fig.~\ref{fig:Ex}, illustrating the excitation shown in panel Ex~2). Only special cases, which significantly disrupt or enhance the periodicity of the localized subsystem, have a potential for crucial influence.

\begin{figure}[!h]
    \centering
    \includegraphics[width=0.9\linewidth]{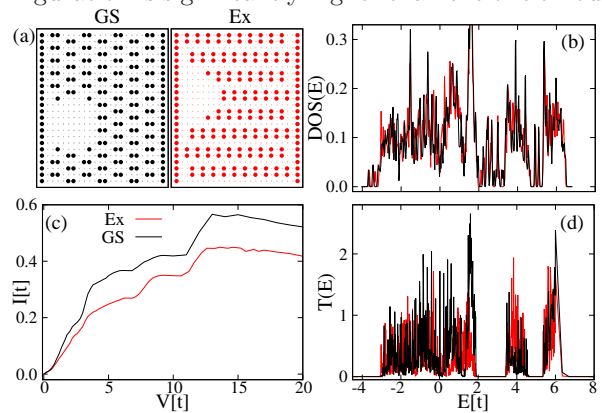}
    \caption{Comparison of transport properties of the ground state configuration (black dots and lines) and excited configuration discussed in the text (red dots and lines). The data were obtained for $n_f=1/3$, $\overline{\mu}=-0.15$ ($n_d\approx1/3$), and $L=24\times 24$. A Gaussian broadening from Eq.~\eqref{eq:DOS} with $\sigma=0.02t$ was used in panel (c).}
    \label{fig:DimEx}
\end{figure} 
The above case described a simple one-particle excitations. For sufficiently wide systems, 
there are also other types of excitations that must be carefully investigated, especially when addressing the 
ground-state properties. One example is shown in  Fig.~\ref{fig:DimEx}. 
The excited configuration in the right panel of Fig.~\ref{fig:DimEx}(a) has higher energy than the ground-state one (the difference is more than $5\%$) shown in the left panel. However, it is an example of a deep local minimum of the energy in the $f$-particle 
configuration landscape. Such a minimum is hard to escape just by local single particle updates. 
This is one of the reasons why we use the climbing algorithm that shifts multiple particles per single update on 
top of the simulated annealing process. 

Interestingly, both configurations from Fig.~\ref{fig:DimEx}(a) lead to comparable transport properties. 
They have an equivalent structure of the transmission functions, shown in Fig.~\ref{fig:DimEx}(d), which 
is also reflected in the $I-V$ characteristics. 

However, the charge current through the ground-state configuration is significantly higher than the one through the excited configuration. This shows the importance of a careful investigation of the annealed ground state and the necessity of averaging the finite-temperature data through a sufficient number of independent MC run sets.

\bibliographystyle{apsrev4-1}
\bibliography{literature}

\end{document}